\documentclass{aastex631}

\usepackage{graphicx}
\usepackage{amssymb}
\usepackage{amsmath}
\usepackage{natbib}
\usepackage{multirow}

\usepackage{xspace}

\newcommand{\swift}{{\it Swift}}

\newcommand{\src}{Swift~J1555.2$-$5402\xspace}
\newcommand{\nicerdays}{29\xspace} 
\newcommand{\nicerlast}{July 1\xspace} 
\newcommand{\distance}{10~kpc\xspace} 
\newcommand{\age}{2.0~kyr\xspace}
\newcommand{\Bsurf}{$3.5\times 10^{14}$~G\xspace}
\newcommand{\Lsd}{$2.1\times 10^{34}$~erg~s$^{-1}$\xspace}
\shorttitle{\textit{NICER} monitoring of the new magnetar \src}
\shortauthors{nicer team}

\begin{document}


\title{A month of monitoring the new magnetar \src during an X-ray outburst}

\correspondingauthor{NICER team}


\author[0000-0003-1244-3100]{Teruaki Enoto}
\affiliation{RIKEN Cluster for Pioneering Research, 2-1 Hirosawa, Wako, Saitama 351-0198, Japan}

\author[0000-0002-0940-6563]{Mason Ng}
\affiliation{MIT Kavli Institute for Astrophysics and Space Research, Massachusetts Institute of Technology, Cambridge, MA 02139, USA}

\author[0000-0001-8551-2002]{Chin-ping Hu}
\affiliation{Department of Physics, National Changhua University of Education, Changhua 50007, Taiwan}

\author[0000-0002-3531-9842]{Tolga G\"uver}
\affiliation{Istanbul University, Science Faculty, Department of Astronomy and Space Sciences, Beyaz\i t, 34119, \.Istanbul, Turkey}
\affiliation{Istanbul University Observatory Research and Application Center, Istanbul University 34119, \.Istanbul Turkey}

\author[0000-0002-6789-2723]{Gaurava K. Jaisawal}
\affiliation{National Space Institute, Technical University of Denmark, Elektrovej 327-328, DK-2800 Lyngby, Denmark}

\author[0000-0002-9700-0036]{Brendan O'Connor}
\affiliation{Department of Physics, The George Washington University, 725 21st Street NW, Washington, DC 20052, USA}
\affiliation{Astronomy, Physics and Statistics Institute of Sciences (APSIS), The George Washington University, Washington, DC 20052, USA}
\affiliation{Department of Astronomy, University of Maryland, College Park, MD 20742-4111, USA}
\affiliation{Astrophysics Science Division, NASA Goddard Space Flight Center, Greenbelt, Maryland 20771, USA}

\author[0000-0002-5274-6790]{Ersin G\"o\u{g}\"u\c{s}}
\affiliation{Sabanc\i~University, Faculty of Engineering and Natural Sciences, \.Istanbul 34956 Turkey}

\author{Amy Lien}
\affiliation{Center for Research and Exploration in Space Science and Technology (CRESST) and
NASA Goddard Space Flight Center, Greenbelt, MD 20771, USA}
\affiliation{Department of Physics, University of Maryland, Baltimore County, 1000 Hilltop Circle,
Baltimore, MD 21250, USA}

\author[0000-0002-2498-1937]{Shota Kisaka}
\affiliation{Department of Physics, Hiroshima University, Higashi-Hiroshima, 739-8526, Japan}

\author[0000-0002-9249-0515]{Zorawar Wadiasingh}
\affil{Astrophysics Science Division, NASA Goddard Space Flight Center, Greenbelt, Maryland 20771, USA}
\affil{Universities Space Research Association (USRA) Columbia, Maryland 21046, USA}

\author[0000-0002-4694-4221]{Walid~A.~Majid}
\affiliation{Jet Propulsion Laboratory, California Institute of Technology, Pasadena, CA 91109, USA}
\affiliation{Division of Physics, Mathematics, and Astronomy, California Institute of Technology, Pasadena, CA 91125, USA}

\author[0000-0002-8912-0732]{Aaron~B.~Pearlman}
\altaffiliation{McGill Space Institute~(MSI) Fellow.}
\affiliation{Department of Physics, McGill University, 3600 rue University, Montréal, QC H3A 2T8, Canada}
\altaffiliation{FRQNT Postdoctoral Fellow.}
\affiliation{McGill Space Institute, McGill University, 3550 rue University, Montréal, QC H3A 2A7, Canada}
\affiliation{Division of Physics, Mathematics, and Astronomy, California Institute of Technology, Pasadena, CA 91125, USA}


\author{Zaven Arzoumanian}
\affiliation{Astrophysics Science Division, NASA Goddard Space Flight Center, Greenbelt, Maryland 20771, USA}

\author[0000-0002-7418-7862]{Karishma Bansal}
\affiliation{Jet Propulsion Laboratory, California Institute of Technology, Pasadena, CA 91109, USA}

\author[0000-0003-4046-884X]{Harsha Blumer}
\affiliation{Department of Physics and Astronomy, West Virginia University, Morgantown, WV 26506, USA}
\affiliation{Center for Gravitational Waves and Cosmology, West Virginia University, Chestnut Ridge Research Building, Morgantown, WV 26505, USA}

\author[0000-0001-8804-8946]{Deepto Chakrabarty}
\affiliation{MIT Kavli Institute for Astrophysics and Space Research, Massachusetts Institute of Technology, Cambridge, MA 02139, USA}

\author{Keith Gendreau}
\affiliation{Astrophysics Science Division, NASA Goddard Space Flight Center, Greenbelt, Maryland 20771, USA}

\author[0000-0002-6089-6836]{Wynn C. G. Ho}
\affiliation{Department of Physics and Astronomy, Haverford College, 370 Lancaster Avenue, Haverford, PA 19041, USA}

\author[0000-0003-1443-593X]{Chryssa Kouveliotou}
\affiliation{Department of Physics, The George Washington University, 725 21st Street NW, Washington, DC 20052, USA}
\affiliation{Astronomy, Physics and Statistics Institute of Sciences (APSIS), The George Washington University, Washington, DC 20052, USA}

\author[0000-0002-5297-5278]{Paul S. Ray}
\affiliation{Space Science Division, U.S. Naval Research Laboratory, Washington, DC 20375, USA}

\author[0000-0001-7681-5845]{Tod E. Strohmayer}
\affiliation{Astrophysics Science Division and Joint Space-Science Institute, NASA's Goddard Space Flight Center, Greenbelt, MD 20771, USA}

\author{George Younes}
\affiliation{Department of Physics, The George Washington University, 725 21st Street NW, Washington, DC 20052, USA}
\affiliation{Astronomy, Physics and Statistics Institute of Sciences (APSIS), The George Washington University, Washington, DC 20052, USA}

\author[0000-0001-7128-0802]{David M. Palmer}
\affiliation{Los Alamos National Laboratory, Los Alamos, NM, 87545, USA}

\author{Takanori Sakamoto}
\affiliation{College of Science and Engineering, Department of Physical Sciences, Aoyama Gakuin University,
5-10-1 Fuchinobe, Chuo-ku,
Sagamihara-shi Kanagawa 252-5258,
Japan}

\author[0000-0001-9399-5331]{Takuya Akahori}
\affiliation{Mizusawa VLBI Observatory, National Astronomical Observatory of Japan, 2-21-1, Osawa, Mitaka, Tokyo 181-8588, Japan}
\affiliation{Operation Division, Square Kilometre Array Observatory, Jodrel Bank Observatory, Lower Withington, Macclesfield, Cheshire SK11 9FT, UK}

\author[0000-0003-0340-0651]{Sujin Eie}
\affiliation{Department of Astronomy, Graduate School of Science, The University of Tokyo, 7-3-1 Hongo, Bunkyo-ku, Tokyo 113-0033, Japan}
\affiliation{Mizusawa VLBI Observatory, National Astronomical Observatory of Japan, 2-21-1, Osawa, Mitaka, Tokyo 181-8588, Japan}





\begin{abstract} 
The soft gamma-ray repeater \src was discovered by means of a 12-ms duration short burst detected with {\it Swift} BAT on 2021 June 3. Then 1.6 hours after the first burst detection, {\it NICER} started daily monitoring of this X-ray source for a month. The absorbed 2--10 keV flux stays nearly constant at around $4\times 10^{-11}$~erg~s$^{-1}$~cm$^{-2}$ during the monitoring timespan, showing only a slight gradual decline. 
A 3.86-s periodicity is detected, and the time derivative of this period is measured to be $3.05(7)\times 10^{-11}$~s~s$^{-1}$. 
The soft X-ray pulse shows a single sinusoidal shape with a root-mean-square pulsed fraction that increases as a function of  energy from 15\% at 1.5~keV to 39\% at 7~keV.  
The equatorial surface magnetic field, characteristic age, and spin-down luminosity are derived under the dipole field approximation to be \Bsurf, \age, and \Lsd, respectively. 
An absorbed blackbody with a temperature of 1.1~keV approximates the soft X-ray spectrum. 
Assuming a source distance of \distance, the peak X-ray luminosity is $\sim 8.5\times 10^{35}$~erg~s$^{-1}$ in the 2--10~keV band. 
We detect 5 and 37 short bursts with \textit{Swift}/BAT and \textit{NICER}, respectively. 
Based on these observational properties, this new source is classified as a magnetar. 
We also coordinated hard X-ray and radio observations with \textit{NuSTAR}, DSN, and VERA. 
A hard X-ray power-law component that extends up to at least 40~keV is detected at 3$\sigma$ significance.
The 10--60~keV flux, which is dominated by the power-law component, is $\sim 9\times 10^{-12}$~erg~s$^{-1}$~cm$^{-2}$ with a photon index of $\sim 1.2$. 
The pulsed fraction has a sharp cutoff above 10~keV, down to $\sim$10\% in the hard-tail component band. 
No radio pulsations were detected during the DSN nor VERA observations. We place 7$\sigma$ upper limits of 0.043\,mJy and 0.026\,mJy on the flux density at S-band and X-band, respectively. 
\end{abstract} 

\keywords{
\href{http://astrothesaurus.org/uat/992}{Magnetars (922)},
\href{http://astrothesaurus.org/uat/994}{Magnetic fields (994)},
\href{http://astrothesaurus.org/uat/1108}{Neutron stars (1108)},
\href{http://astrothesaurus.org/uat/1306}{Pulsars (1306)},
\href{http://astrothesaurus.org/uat/1471}{Soft gamma-ray repeaters (1471)},
\href{http://astrothesaurus.org/uat/1852}{X-ray transient sources (1852)},
}



\section{Introduction} 
\label{sec:Introduction}
Magnetars are highly-magnetized neutron stars that are usually bright in X-rays as a result of the release of an enormous amount of magnetic energy stored in the stellar interior and the magnetosphere \citep{2008A&ARv..15..225M,2017ARA&A..55..261K}. 
Among them, sources emitting repetitive soft gamma-ray bursts are historically called soft gamma-ray repeaters (SGRs, e.g., \citealt{1998Natur.393..235K}). 
In the last decade, systematic monitoring of magnetars in the X-rays, mainly with X-Ray Telescope (XRT) onboard the Neil Gehrels {\it Swift} Observatory, revealed that many transient magnetars spend most of their time in a quiescent state with low activity. However, they occasionally exhibit a sudden X-ray brightening where the X-ray flux reaches an initial plateau  $10^{-11}$--$10^{-10}$\,erg\,s$^{-1}$\,cm$^{-2}$ lasting a few weeks, followed by the gradual decay over a couple of months \citep{2017ApJS..231....8E,2018MNRAS.474..961C}. 
These magnetar outbursts are characterized by enhanced persistent X-ray emission, sporadic short bursts, pulsar timing anomalies, and, rarely, giant flares. 
Their origin has been attributed to various mechanisms, such as the relaxation process of twisted magnetic fields, starquakes, magnetothermal evolution, and magnetic field dissipation  \citep{1995MNRAS.275..255T,1996ApJ...473..322T,2011ApJ...727L..51P,2016ApJ...833..261B}. Multi-wavelength observations of these outbursts are essential to address a wide range of astronomical topics, as demonstrated, for example, with the discovery of Fast Radio Bursts (FRBs) associated with a short hard X-ray burst from the Galactic magnetar SGR~J1935+2154 in 2020 \citep{2020Natur.587...54C,2020Natur.587...59B,2020ApJ...898L..29M,2021NatAs...5..372R,2021NatAs...5..401T,2021NatAs.tmp...54L}. Such a connection between magnetars and FRBs are also supported by the indication that  extragalactic FRBs have statistical signature of magnetar short bursts (e.g.\citealt{2019ApJ...879....4W}).

On 2021 June 3, a new SGR, \src, was discovered through a short burst detection with the Burst Alert Telescope (BAT) onboard {\it Swift}  \citep{2021GCN.30120....1P}. Immediately after the notification of the burst from this source, several X-ray satellites started follow-up observations of this magnetar candidate. These observations were promptly used to measure the spin frequency and frequency derivative \citep{2021ATel14674....1C,2021ATel14684....1N,2021ATel14685....1I} and detected several short bursts \citep{2021ATel14690....1P}. 
Based on the measured strong magnetic field and its distinctive magnetar characteristics, this new source was classified as a magnetar.
In addition, several radio telescopes searched for radio emission and pulsations \citep{2021ATel14678....1B,2021ATel14680....1B,tmp2021ATel14799}. In this \textit{Letter}, we report on the X-ray temporal and spectral characteristics of this new magnetar observed with \textit{Swift}, the Neutron star Interior Composition Explorer (\textit{NICER}), and the Nuclear Spectroscopic Telescope Array (\textit{NuSTAR}) during the initial \nicerdays days of its X-ray outburst; our observations were also coordinated with radio monitoring. 
Here we adopt a fiducial distance $d$ of \src at \distance and the normalization factor $d_{10}=d/(10\,\textrm{kpc})$ (see discussion \S\ref{sec:associations}).

\section{Observation and Data Reduction}
\label{sec:Observation and Data Reduction}

\subsection{Swift}
The {\it Swift} BAT \citep{2004ApJ...611.1005G} detected a burst from an unknown source at 09:45:46 UT on 2021 June 3 (trigger number 1053220), and immediately pointed to the source direction \citep{2021GCN.30120....1P}. 
The {\it Swift} XRT \citep{2005SSRv..120..165B}
obtained X-ray data in the WT mode for 62 s from 97 s after the BAT trigger and then in the PC mode for $\sim 1.7$ ks from 1.1 hr after the burst.
The XRT observations determined the source position (J2000.0) to be R.A.$=15^{\textrm h}\,55^{\textrm m}\,08.66^{\textrm s}$ and Decl.$=-54^{\circ}\,03\arcmin\,41.1\arcsec$ with an uncertainty of 2.2\arcsec\ radius at  90\% confidence level \citep{2021ATel14675....1E}. We adopted this source position for all analyses presented in this {\it Letter}. The BAT detected another four short bursts from the same direction, as summarized in Appendix Table~\ref{tab_burst_bat}.
We analyzed the BAT data using the standard HEASoft BAT pipelines (version 6.28), following the same procedure described in \citet{Lien16}; these results are present in \S\ref{Short Bursts}.

We analyzed the XRT data obtained on June 3, 4, 5, and 7.
The observation IDs (ObsIDs) used in this {\it Letter} are listed in Appendix Table \ref{tab_obsid_swift}. The observations on June 4, 5, and 7 were carried out in the WT mode for a total of 8.9 ks. We processed the data through the standard procedure of FTOOLS \texttt{xrtpipeline} with the default filtering criteria and extracted source photons from a circular region with a 20-pixel radius (1 pixel\,$=$\,$2.36\arcsec$) centered at the target, whereas we collected background spectra from a source-free region with a similar (20-pixel) radius, located far ($>$\,$2\arcmin$) from the source. We used the latest available RMF file in CALDB version 20210504. We generated the ARF files with the \texttt{xrtmkarf} tool. 

\subsection{NICER}
{\it NICER} (\citealt{2016SPIE.9905E..1HG}) onboard the International Space Station (ISS) began X-ray observations of the source at 11:21:31 UT on 2021 June 3, 1.6 hours after the first short burst detected with \textit{Swift} BAT. This initial {\it NICER} observation was  2.4~ks long in exposure (ObsID 4202190101), and it was followed by high-cadence monitoring (see Appendix Table~\ref{tab_obsid_nicer}) carried out almost daily for \nicerdays ~days under an approved cycle 3 proposal.
Each ObsID had roughly 2~ks exposure and was divided into several continuous good time intervals (GTIs) with exposures of a few hundred seconds for each. 

The \textit{NICER}'s X-ray Timing Instrument (XTI) has on-orbit 52 active modules, each of which consists of co-aligned X-ray concentrators and silicon drift detectors. The XTI has a time resolution of $<$100 ns, and the total effective area is about 1,800~cm$^2$ at around 1.5 keV. We performed the standard analysis procedures using \texttt{NICERDAS} (version 2020-04-23\_V007a) in HEASoft 6.27.2 and NICER calibration database (version 20200722).
We generated level-2 cleaned events with the \texttt{nicerl2} command. 
For the barycentric correction, we used  \texttt{barycorr} with Jet Propulsion Laboratory Solar system development ephemeris DE405 for the source coordinates stated above \citep{standish98}.
For timing analyses and burst searches, we utilized all active 52 modules.
For spectral studies, we further excluded module numbers 14 and 34 to avoid potential contamination by instrumental noise in the soft energy band. The background spectral model is generated using the 3C50 background model with \texttt{nibackgen3c50} command \citep{2021arXiv210509901R}\footnote{\url{https://heasarc.gsfc.nasa.gov/docs/nicer/tools/nicer_bkg_est_tools.html}}.

\subsection{NuSTAR}
{\it NuSTAR} (\citealt{2013ApJ...770..103H}) observed \src on June 5--6 for 38.4~ks exposure and two additional contemporaneous observations with \textit{NICER} on June 9 (25~ks exposure) and 21 (29~ks). 
We processed and filtered the \textit{NuSTAR} data following the standard procedures with HEASoft version 6.28 and CALDB version 20210524, using the \texttt{nupipline} and \texttt{nuproducts} commands. We extracted on-source and background spectra from circular regions of 80$''$-radius centered at the source position and in a source-free region, respectively.  
The background-subtracted source count rates of FPMA was about 1 count~s$^{-1}$ in the 3--79~keV band. 
For the spectral fitting, we grouped source spectra using the \texttt{grppha} tool of HEASoft, such that each spectral bin would have a minimum of 50 counts. 

\subsection{Deep Space Network (DSN)}
We carried out radio observations of \src for a total exposure of roughly 10.8 hours at five epochs during 2021 June 4--12 using different Deep Space Network (DSN) radio telescopes (see Table~\ref{tab_obsid_dsn}).  Simultaneous dual-frequency bands, with center frequencies at 2.2 GHz (S-band) and 8.4 GHz (X-band), were used  for all observations. We used a single circular polarization mode for DSS-34 and DSS-36, whereas a dual circular polarization mode was used at each frequency band for DSS-43.

We recorded the data in filterbank mode with a time resolution of 512~$\mu$s and frequency resolution of 1\,MHz using the pulsar machine in Canberra.  The data processing procedure follows similar steps to those presented in earlier studies of pulsars and magnetars with the DSN (e.g.,~\citealt{2017ApJ...834L...2M, 2018ApJ...866..160P, 2019AdAst2019E..23P}).  After first flattening the bandpass response in each data set, we removed the low-frequency variations in the temporal baseline of each frequency channel by subtracting the moving average from each data point with a time constant of 10 seconds. The sample times were then corrected to the solar system barycenter.   

We dedispersed the data of each epoch with trial DMs between 0 and 5000~$\mathrm{pc\,cm^{-3}}$ and subsequently searched each resulting time series for both periodic and single pulse emission.  We found no statistically significant periods with a signal-to-noise ratio (S/N) above 7.0 after folding individual dedispersed time series modulo period candidates from PRESTO’s \texttt{accelsearch} package. In addition, we folded the dedispersed time series at each DM trial using the timing model from NICER in Table \ref{tab:summary_parameter}, but found no evidence of radio pulsations at S-band or X-band during any of our observations. For each epoch, we place 7$\sigma$ upper limits on the magnetar's flux density, assuming a duty cycle of 10\% (see Table~\ref{tab_obsid_dsn}). Based on our longest observation with DSS-43, the 70\,m radio telescope in Tidbinbilla, Australia, we obtain 7$\sigma$ upper limits on the magnetar's flux of $<$\,0.043\,mJy at S-band and $<$\,0.026\,mJy at X-band. 

We also searched the dedispersed time series at each frequency band for radio bursts using a matched filtering algorithm, where the time series was convolved with boxcar functions with logarithmically spaced widths between 512~$\mu$s and 150~ms.  Candidates with a detection S/N above 7.0 were saved and classified using the FETCH software package~\citep{2020MNRAS.497.1661A}.  The dynamic spectra of the candidates were also visually inspected for verification.  
We detected no radio bursts during the radio observations and place 7$\sigma$ upper limits on the fluence of individual bursts during each epoch at both S-band and X-band (see~Table~\ref{tab_obsid_dsn}).
On June 5, 2021, we detected an X-ray burst, with a width of $w$\,$=$\,15.91\,ms (see Table~\ref{tab_burst_nicer}; burst \#8), during an overlapping radio and X-ray observation. 
However, no prompt radio emission (within $\pm$10\,s of the X-ray burst time) was detected above a 7$\sigma$ fluence detection threshold of 1.6$\sqrt{w/\text{1\,ms}}$\,Jy\,ms and 0.61$\sqrt{w/\text{1\,ms}}$\,Jy\,ms at S-band and X-band, respectively.

\subsection{VERA (K-band) \label{sec:vera}}
The 20-m-diameter Ishigaki-jima station of VLBI Exploration of Radio Astrometry (VERA) conducted a ToO observation of this source at an observation frequency of 22 GHz (1.3 cm, K-band) with a bandwidth of 512 MHz. The acquired data for one hour at 14:40--15:40 UT on 2021 June 6 were processed and folded to explore radio pulsations both with and without assuming the rotation period. 
Because the data quality was limited due to a low elevation of the object and bad weather conditions, we could only set an upper limit of the peak flux density of 1.02 Jy (1$\sigma$). 

\section{Analysis and Results}
\label{sec:Data Analysis and Results}

\subsection{X-ray timing analyses \label{sec:timing}} 
Figure~\ref{fig:monitoring} shows the time-series of physical parameters of \src during {\it NICER} monitoring for the \nicerdays days from shortly after the onset of the outburst until \nicerlast.
In constructing the time-series, we first derived the pulsar spin ephemeris, for which we used a Gaussian pulse template and constructed pulse times of arrival (TOA) with an integration time of 300 s and a minimum exposure of 200 s contained in each bin, using the script \texttt{photon\_toa.py} from \texttt{NICERsoft}\footnote{\url{https://github.com/paulray/NICERsoft/}}. 
The timing analysis was carried out over 2--8 keV, where the energy range was determined from a $Z_n^2$ search with $n=2$ to optimize the pulse significance \citep{1983A&A...128..245B}. 
We used the Python-based package for high-precision timing analysis ``PINT" \citep[version 0.8.2]{2021ApJ...911...45L} to compute the best timing model through a weighted least-squares fit to the TOAs.
The TOAs were found to be well described by either a fifth-order polynomial model, as summarized in Table~\ref{tab:summary_parameter}, or a glitch model with three glitch candidates (see Appendix \ref{sec:appendixtiming}) for the spin evolution of \src. 
The best-fit frequency and its derivatives are 
$\nu = 0.258997103(8)$, $\dot{\nu} = -2.04(5)\times10^{-12}$ Hz/s, and $\ddot{\nu} = -4.50(13)\times10^{-18}$ Hz/s${^2}$, $\nu^{(3)} = -1.10(10)\times10^{-23}$ Hz s$^{-3}$, $\nu^{(4)} = 3.59(15)\times10^{-29}$ Hz s$^{-4}$, and $\nu^{(5)} = 1.59(14)\times10^{-34}$ Hz s$^{-5}$ at barycentric epoch $T_0 = {\rm MJD}\ 59382.7549$. 
{\it NICER}'s sensitivity enables our measurement up to fifth-order in frequency with TOAs of only 300~s.

In Figure~\ref{fig:pulse_profile} a--d, we present the energy-resolved and background-subtracted pulse profiles of \src in the 2--3 keV and 3--8 keV bands with \textit{NICER} and in the 3--8 keV, 8--12 keV, and 12--20 keV with \textit{NuSTAR}, where estimates of the background rates were made with the \texttt{nibackgen3C50} tool for the \textit{NICER} data and were based on the measured rate in the background region for the \textit{NuSTAR} data. The soft X-ray profile (2--12\,keV) shows a single-peaked, nearly sinusoidal shape, 
while the pulsation in the hard X-ray band ($\gtrsim$12\,keV) was hardly detected. 
We further divided the data into finer energy bands and calculated, in Figure \ref{fig:pulse_profile}f, the time-averaged root mean square (RMS) pulsed fraction (PF) as defined in \citet{1997ApJS..113..367B,2004ApJ...605..378W}. 
Both the \textit{NICER} and \textit{NuSTAR} observations suggest that the
RMS PF 
increases from $\sim30\%$ in the softer band (3--4 keV) to the maximum of $\sim$40\% at around 7 keV, and then decreases with energy to $\lesssim$20\%. 
We also found that the PF 
in the 3--8 keV range remained almost constant during the observed period (Figure~\ref{fig:monitoring}e).

\subsection{X-ray spectral analyses
\label{sec:spectrum}}

\begin{figure*}[h]
\begin{center}
\vspace{2mm}
\includegraphics[width=0.48\textwidth]{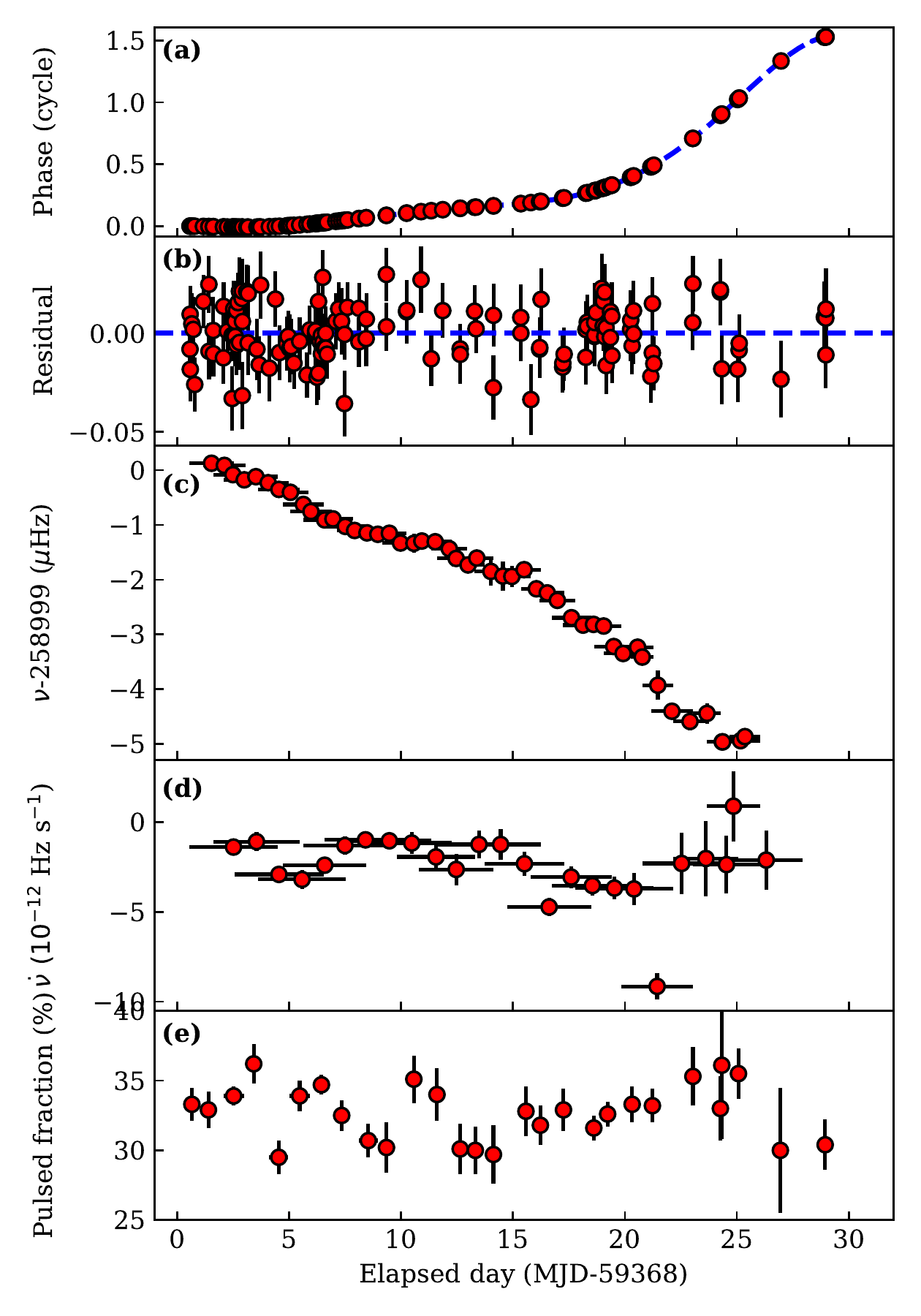}
\includegraphics[width=0.48\textwidth]{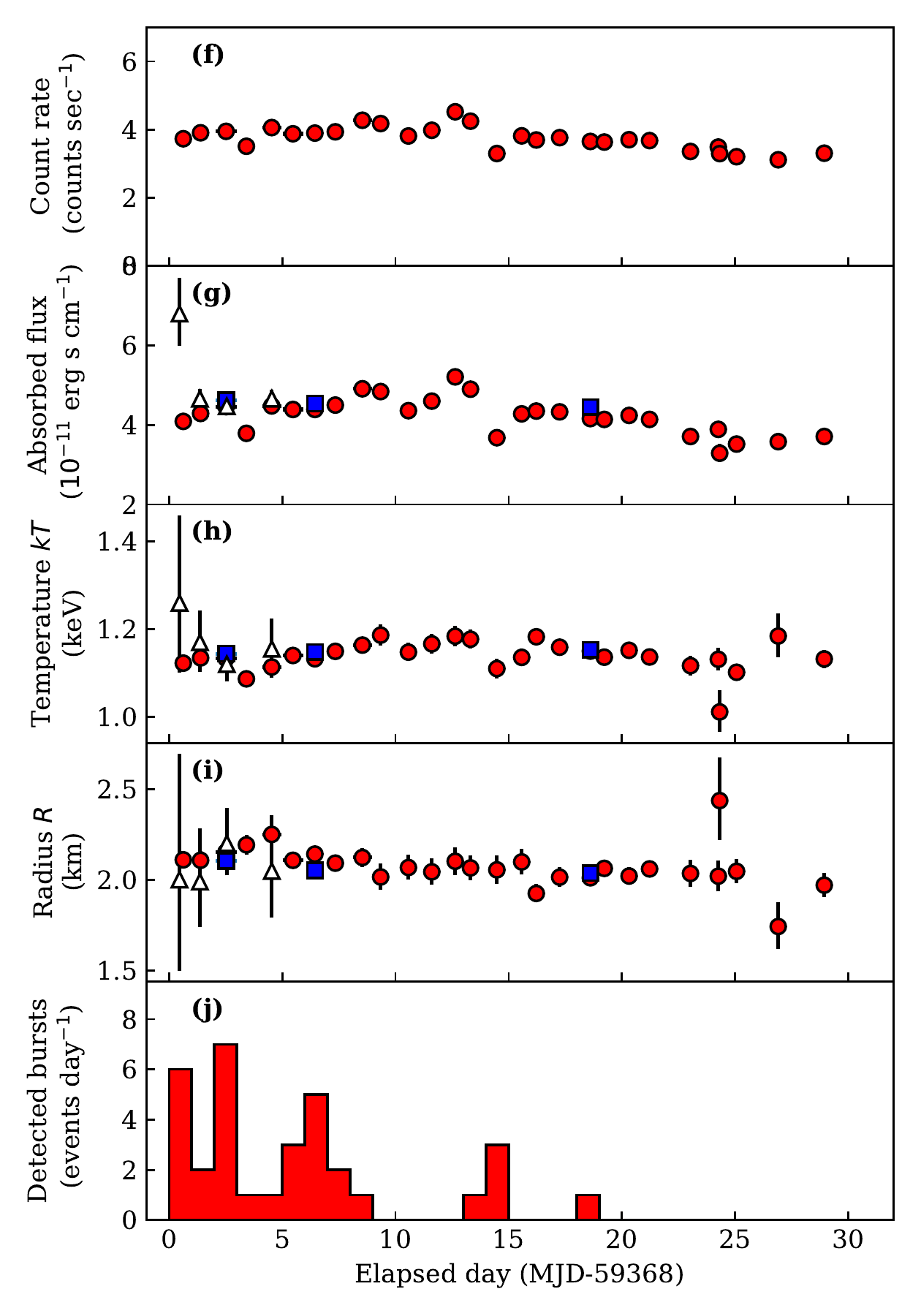}
\caption{
{\it NICER} monitoring of the timing (left panels) and spectral evolutions (right panels) of the 2021 outburst of \src  over  \nicerdays ~days (2021 June 3--\nicerlast; MJD 59368--59396).
The time origin MJD 59368 is the day when the first burst was detected with \textit{Swift}/BAT.
(a) Intrinsic pulse-phase evolution  with respect to a folding frequency of $\nu_{\rm fold} = 0.258997274$ Hz and a folding frequency derivative of $\dot{\nu}_{\rm fold} = -1.63\times10^{-12}$ Hz s$^{-1}$; dashed line is the best-fit model with a fifth-order polynomial.    
(b) Phase residuals (cycles) after correcting for the spin derivatives (up to 5th order).
(c) Spin frequency with 2-day windows in steps of 0.5 days.
(d) Spin frequency derivative with 4-day windows in steps of 1 day.
(e) RMS pulsed fraction in the 3--8~keV band.
(f) Background-subtracted 2--10\,keV \textit{NICER} count rate.
(g) 2--10 keV absorbed X-ray flux obtained with
\textit{Swift} (open triangles),
\textit{NICER} only (red circles), and \textit{NICER} simultaneously fitted with   \textit{NuSTAR} (blue squares). The symbols are the same in panels h and i. 
(h) Blackbody temperature (keV). 
(i) Emission radius of the blackbody component assuming a fiducial distance of \distance.
(j) Number of short bursts per day detected with \textit{NICER} and \textit{Swift}/BAT.
Error bars are 68\% confidence limit in these plots. 
\label{fig:monitoring}}
\vspace{-0.4cm}
 \end{center}
\end{figure*}

\begin{figure*}[h]
\begin{center}
\vspace{2mm}
\includegraphics[width=0.48\textwidth]{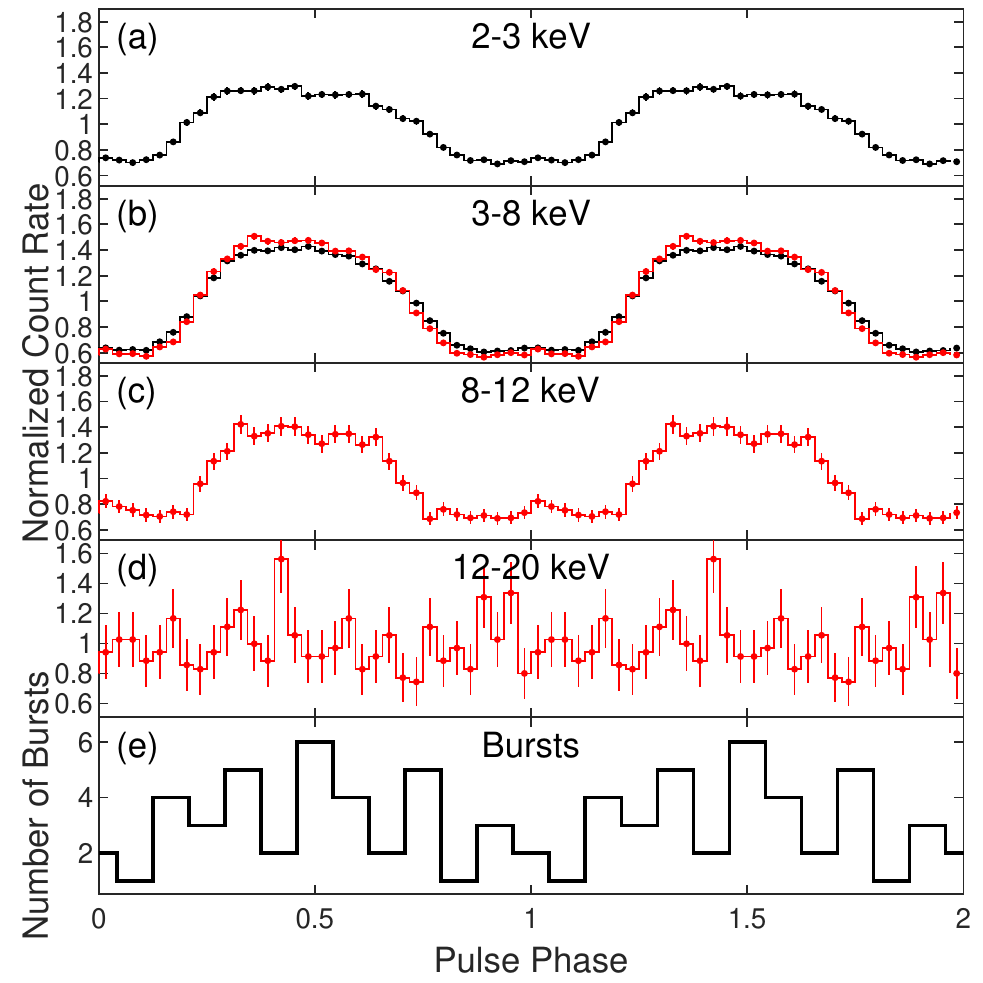}
\includegraphics[width=0.48\textwidth]{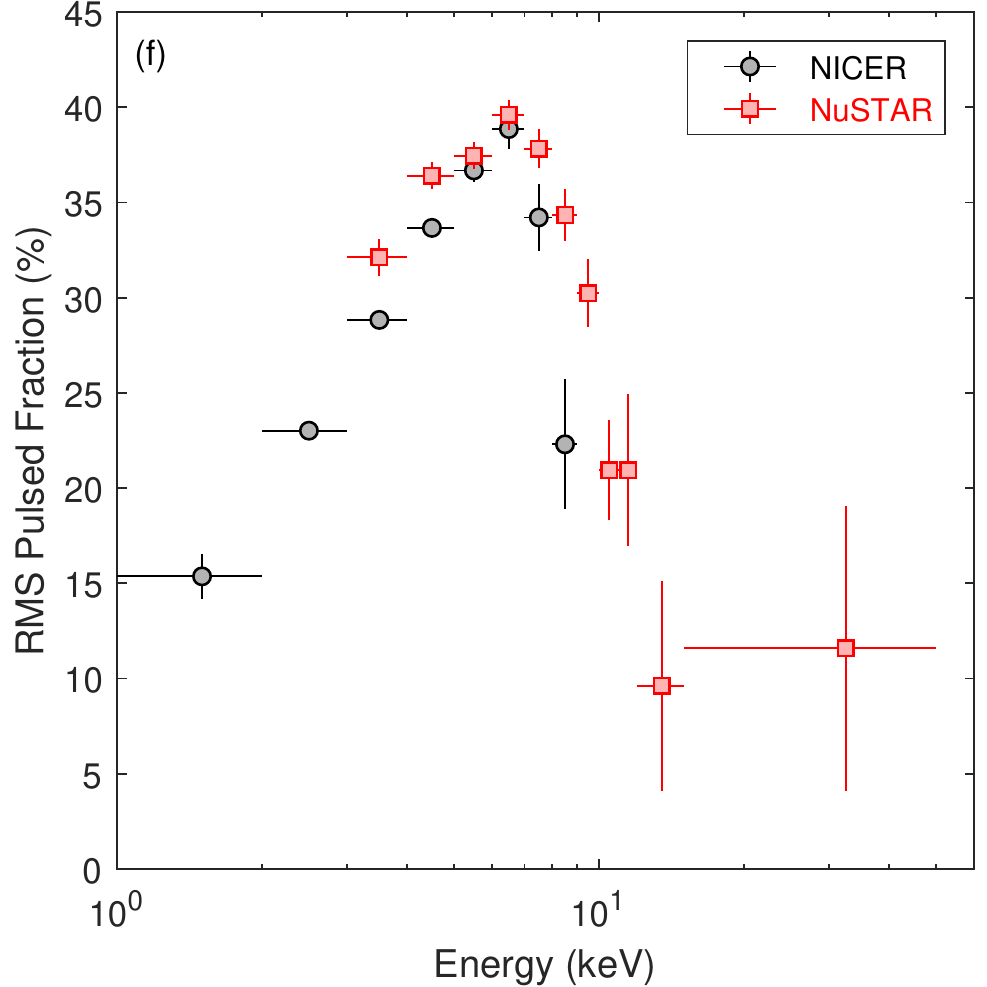}
\caption{
Left: Panels (a)--(d) are background subtracted X-ray pulse profiles of \src in the 2--3~keV, 3--8~keV, 8--12~keV, and 12--20~keV, respectively, taken with (black) \textit{NICER} and (red) \textit{NuSTAR}. 
The amplitudes are normalized relative to the mean count rate. 
Two cycles are shown in this figure for clarity. 
Error bars indicate 1$\sigma$ uncertainties. 
Panel (e) shows the phase distribution of short bursts.
Right: RMS pulsed fraction as a function of energy. \label{fig:pulse_profile}}
\vspace{-0.4cm}
 \end{center}
\end{figure*}

The right panels of Figure~\ref{fig:monitoring} show the long-term spectral properties of \src obtained with \textit{Swift}, \textit{NICER}, and \textit{NuSTAR} data. 
Here we applied a single-temperature blackbody multiplied by the Tuebingen-Boulder interstellar absorption model (\texttt{tbabs*bbodyrad} in \texttt{Xspec} terminology) to derive the physical parameters at each epoch. The first data point in all right panels corresponds to the initial 1.7-ks {\it Swift} PC spectrum obtained $\sim$1.1 hours after the BAT trigger, which is well fitted with the model above (chi-square of 113 for 128 degrees of freedom; dof). We derived the best-fit hydrogen column density of $N_{\rm H}=6.8^{+2.0}_{-1.7}\times 10^{22}$~cm$^{-2}$ and blackbody temperature of $kT=1.26^{+0.20}_{-0.16}$~keV. The absorbed 2--10 keV flux is  $6.8^{+0.9}_{-0.8}\times 10^{-11}$ erg cm$^{-2}$ s$^{-1}$. 
This flux is significantly higher than those in the following observations with the \textit{Swift}/XRT in the WT mode and \textit{NICER} monitoring.

The subsequent daily \textit{NICER} spectra (1.7--10\,keV) were systematically fitted with the same model with the hydrogen column density tied to be the same value among  all \textit{NICER} spectra at $N_{\rm H} =(8.88\pm0.12)\times 10^{22}$~cm$^{-2}$. Each observation has $\sim$4 counts~sec$^{-1}$ (Figure~\ref{fig:monitoring}f).
The reduced chi-square values were $\sim$0.9--1.2 for $\sim$100-400~dof. 
No spectral variation during the initial monitoring was found. The absorbed and unabsorbed 2--10 keV fluxes were $\sim 4.3\times 10^{-11}$~erg~cm$^{-2}$~s$^{-1}$ (Figure~\ref{fig:monitoring}g) and $\sim 7.5\times 10^{-11}$~erg~cm$^{-2}$~s$^{-1}$, respectively. The derived temperature and emission radius were constant at $\sim$1.1~keV  (Figure~\ref{fig:monitoring}h) and $\sim 2\,d_{10}$~km  (Figure~\ref{fig:monitoring}i), respectively (\S\ref{sec:associations}). 
These \textit{NICER} parameters are consistent with those obtained with the WT mode data of \textit{Swift}/XRT. 

\begin{figure*}[h]
\begin{center}
\vspace{2mm}
\includegraphics[width=0.48\textwidth]{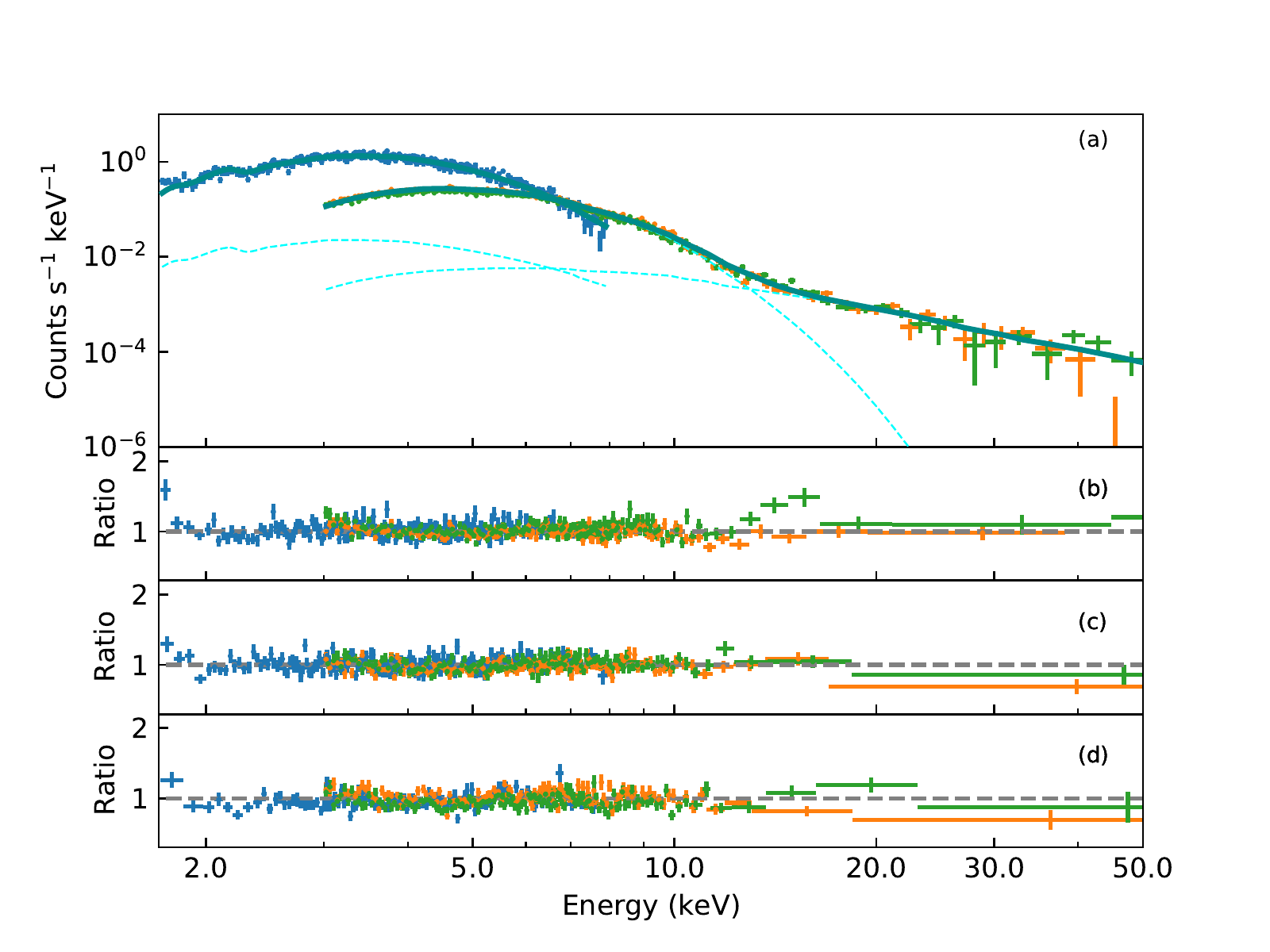}
\includegraphics[width=0.48\textwidth]{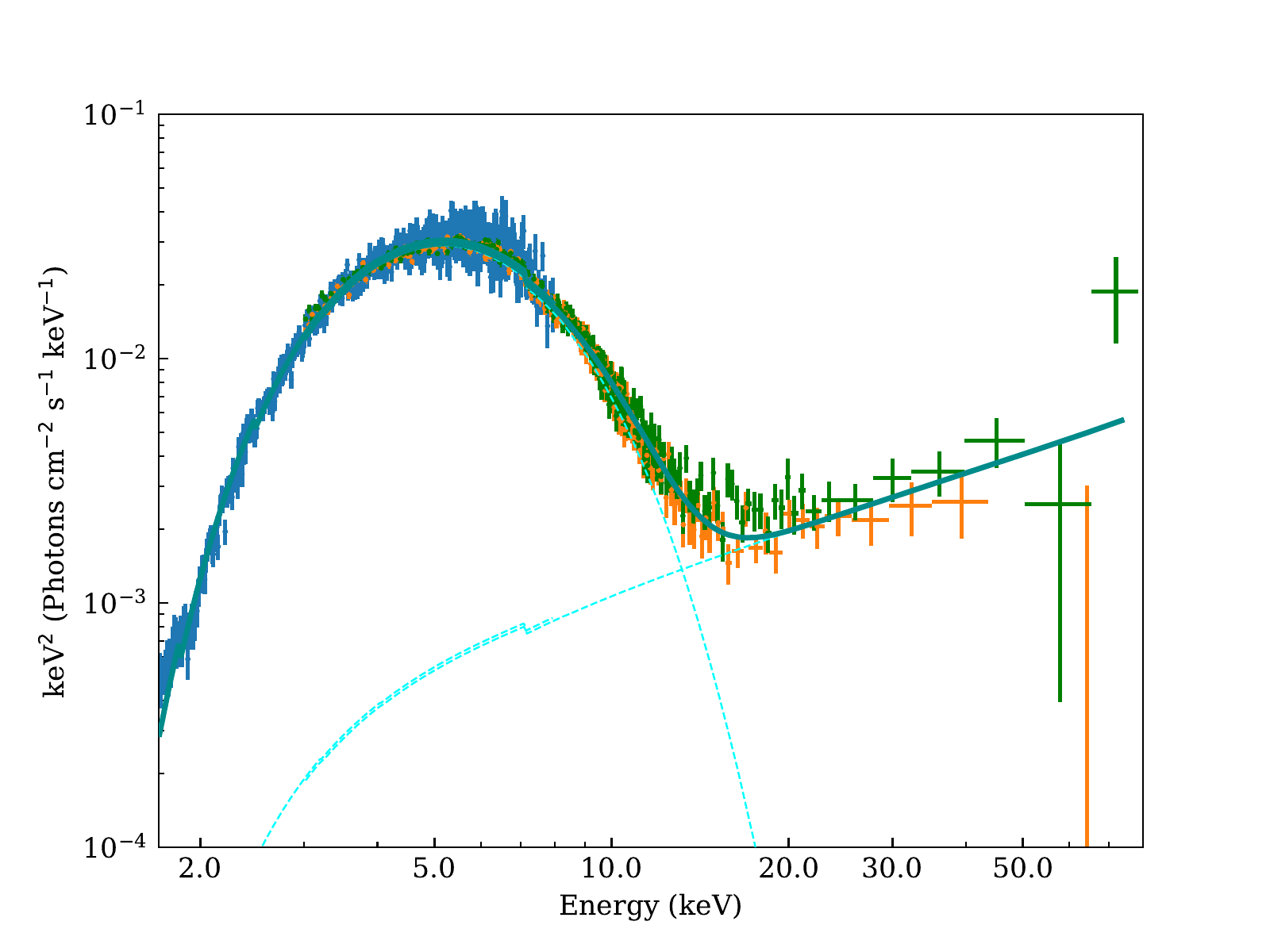}
\caption{
Left: Spectral fitting of the joint {\it NICER} and {\it NuSTAR} data of \src. 
Panel (a) shows the background-subtracted  response-inclusive spectra obtained on June 5 and the best-fit model (dark cyan solid line) with its blackbody and power-law components (cyan dashed lines). Photoelectric absorption is not corrected. Lower panels (b)-(d) show spectra obtained on June 5, 9, and 21, divided by the best-fit model to the first epoch shown in panel (a). 
Right: 
Best-fit $\nu F_{\nu}$ spectra of {\it NICER}, {\it NuSTAR} FPMA and FPMB for the three epochs combined. 
In both panels, {\it NICER} and {\it NuSTAR} FPMA and FPMB data are shown in blue, orange, and green, respectively. 
\label{fig:tmp_fig_xray_spectrum}}
\vspace{-0.4cm}
 \end{center}
\end{figure*}

We performed joint spectral fits of three observations of \textit{NICER} and \textit{NuSTAR} on June 5--6, 9, and 21. 
Since the \textit{NICER} spectra showed no significant time variation, we extracted the \textit{NICER} spectra for the period on the same day as of \textit{NuSTAR} and regarded them as simultaneous even if their observation periods were not fully simultaneous. 
The column density among the three epochs are tied to the same value.
The best-fit spectral model is shown in Figure~\ref{fig:tmp_fig_xray_spectrum}. In addition to the soft X-ray blackbody component, a hard X-ray component above 10\,keV was detected with 3$\sigma$ significance extending up to at least 40\,keV. The hard X-ray flux, when fitted by the power-law model, was $(7-9)\times 10^{-12}$ erg cm$^{-2}$ s$^{-1}$ in the 10--60 keV band with a power-law photon index of $1.2$--$1.7$. We also performed a combined fit of all the three epochs, given that no significant spectral change was observed between them, except for the hard X-ray flux, which  showed a slight decline. 
The resultant average $\nu F_{\nu}$ is shown in the right panel of  Figure~\ref{fig:tmp_fig_xray_spectrum} right and Table~\ref{tab:summary_parameter}. 
The hard X-ray component is distinctive from the soft blackbody emission below 10\,keV. 

\subsection{Short burst analyses} 
\label{Short Bursts}
\textit{Swift} BAT detected 5 short bursts, as summarized in Appendix Table~\ref{tab_burst_bat}.
For the first detected burst on June 3 (an onboard trigger), we used the data from $T-240$\,s to $T+100$\,s, where $T$ is the burst detection time, whereas we used the $\sim 3$-s interval events collected through sub-threshold triggers\footnote{These are also called failed triggers, which are detections that pass the rate trigger criteria but failed the image detection threshold.} in our analyses of the other bursts. The BAT event data have time resolution of $\sim 100$\,$  \mu$s \citep{Barthelmy05}. The BAT temporal analysis utilizes light curves binned in 1, 2, and 4 ms. 
Our BAT spectral analysis was performed using spectra created with the $T_{90}$ duration of each burst, which is the duration that covers 90\% of the burst emission.
All the spectra were successfully fitted with a single blackbody model (\texttt{bbodyrad} model in \texttt{Xspec}) except for the one on June 7, the statistics of which were too poor to give meaningful constraints.

We also searched the 2--8 keV \textit{NICER} event data for short bursts using the Bayesian block technique\footnote{\href{https://docs.astropy.org/en/stable/api/astropy.stats.bayesian_blocks.html}{https://docs.astropy.org/en/stable/api/astropy.stats.bayesian\_blocks.html}} \citep{ScargleNJ2013}. 
The blocks with high backgrounds and with durations longer than $\sim$1\,s are further filtered out on the basis of comparison between the house-keeping data (\texttt{mkf} files), multiple blocks in one burst, and blocks close to the GTI boundaries. 
We used the Poisson probability to determine the significance of detecting a number of photons in a block, where the non-burst count rate was calculated from  1~s intervals close in time to the bursts. We identified 37 short bursts exceeding $5\sigma$ detection significance as summarized in Appendix Table~\ref{tab_burst_nicer}. 
The average duration of the bursts was $23\pm 17$\,ms and 13 
photons were detected in a burst on average. 
Note that burst 26-1 among the list occurred during the tail of burst 26 and  we included it in burst 26 in the calculation.

We stacked the detected bursts to obtain an average spectrum and found it to be equally well fitted with both a single blackbody (\texttt{tbabs * bbodyrad}) and with a power-law (\texttt{tbabs * pegpwrlw}) with  Cash statistics of $C\textrm{-stat}=256.3$ and $C\textrm{-stat}=259.6$, respectively, with 309 dof. When fixing the absorption column density at $N_{\rm H}=8.72\times 10^{22}$~cm$^{-2}$ (Table~\ref{tab:summary_parameter}), 
the former model gave a blackbody temperature of $2.8_{-0.5}^{+0.7}$ keV, whereas the latter gave a photon index of $0.0\pm0.2$. 
Using the blackbody model, we find an average unabsorbed flux of $(1.3\pm0.1)\times10^{-8}$ ergs s$^{-1}$ cm$^{-2}$ in the 2--8 keV range. 
Assuming a distance of \distance, the blackbody radius is estimated to be $7.4\pm1.6$ km. 
The fluences of detected bursts are calculated to be in the range of (1--13)$\times10^{-10}$~ergs~cm$^{-2}$ with an assumed blackbody spectrum of $kT=2.8$ keV using the WebPIMMs Appendix (Appendix Table~\ref{tab_burst_nicer}). One of the \textit{NICER} bursts was simultaneously detected with \textit{Swift} BAT (Appendix Figure~\ref{fig:BAT_bursts}).

\section{Discussion and Conclusion}
\label{sec:Discussion}

\subsection{Timing and spectral characteristics of the new magnetar}
High-cadence monitoring with \textit{NICER} over one month allows us to measure the spin ephemeris of the new source \src (Table \ref{tab:summary_parameter}). Refined from the initial reports in GCNs and ATels \citep{2021ATel14674....1C, 2021ATel14684....1N}, the period and its time derivative are measured to be 3.86104705(12) sec and $3.05(7)\times 10^{-11}$~s~s$^{-1}$, respectively. 
The combination of the two values falls 
within the distribution of known magnetars on the $P$-$\dot{P}$ diagram (Figure \ref{fig:discussion}a). Assuming the standard rotating magnetic dipole model and a braking index of $n=3$, these timing parameters correspond to a characteristic age of $\tau = 2.01(5)$ kyr, surface magnetic field strength of $B_{\rm surf} = 3.47(4)\times10^{14}$ G, and spin-down luminosity of $L_{\rm sd} = 2.09(5)\times10^{34}$ erg~s$^{-1}$. 
This source was classified as a magnetar based on the measured strong magnetic field. 
The derived characteristic age suggests that it is one of the youngest magnetars among the known ones.
The suggestion is supported by the observed strong timing noise, which requires a model with high-order polynomials (\S\ref{sec:timing}), similar to that of the young magnetar Swift~J1818.0$-$1607 \citep{2020ApJ...902....1H}.
We caution that 
the derived pulsar parameters during the outburst may deviate from those in the quiescent state (e.g., \citealt{2017ApJ...851...17Y,2020ApJ...889..160A}).
Frequency derivatives are known to fluctuate during magnetar outbursts, with variations of a factor of 1--50 
(see, e.g., \citealt{2012ApJ...748....3D,2014ApJ...784...37D,2019MNRAS.488.5251L}).
Thus, the accuracy of the inferred parameters $B_{\rm surf}\propto \sqrt{\dot{P}}$, $\tau\propto \dot{P}^{-1}$, and $L_{\rm sd}\propto \dot{P}$ relative to the quiescent values still have uncertainties due to this $\dot{P}$ variation. 

We detect a hard X-ray tail above 10 keV with \textit{NuSTAR}, extending up to at least 40\,keV with 3$\sigma$ significance. 
The spectral energy distribution  shows that the hard X-ray component is distinguished from the blackbody component, which should originate from the stellar surface (Figure~\ref{fig:tmp_fig_xray_spectrum}). 
The existence of the distinctive hard tail is further supported by the steep drop in the energy-dependent PF 
above 10\,keV (Figure~\ref{fig:pulse_profile}f). 
Two-component spectra of this kind are reported from other persistently-bright and transient magnetars \citep{2006ApJ...645..556K,2010ApJ...722L.162E,2017ApJ...847...85Y}. The low PF 
($\sim$10\%) of \src in the hard X-rays may suggest that the hard tail originates in magnetospheric emission that does not have much anisotropy and higher emission altitude than emission from the stellar surface. This may imply a low magnetic impact parameter \citep[e.g.,][]{2018ApJ...854...98W} for the observer across the pulse for resonant Compton scattering. The 15--60\,keV flux of \src, $F_{15-60}=6.42_{-0.68}^{+0.14}\times 10^{-12}$~erg~s$^{-1}$~cm$^{-2}$, is lower than the absorbed 1--10~keV flux  $F_{1-10}=4.65_{-0.27}^{+0.19}\times 10^{-11}$ ~erg~s$^{-1}$~cm$^{-2}$. Accordingly, the broadband hardness ratio of the magnetospheric to surface-thermal emissions is $\eta=F_{15-60}/F_{1-10}=0.14$. The hardness ratio $\eta$ of known magnetars ($\eta=0.1$--4) is suggested to be correlated with the surface magnetic field. Figure~\ref{fig:discussion}d plots  $\eta$ values of known sources and \src. It shows that $\eta$ of \src is not largely off, though apparently smaller than, the proposed correlation (see equation 3 of \citealt{2017ApJS..231....8E}).
We note that the measured $\eta$ has some systematic uncertainty.
If the yet unknown long-term frequency derivative of \src is lower than the value measured during the current outburst, the magnetic field strength in the quiescent state is weaker than our estimate, which places $\eta$ of \src closer to the known correlation.

\subsection{Search for a counterpart} 
\label{sec:associations}

We searched the \textit{Swift} archival data for a serendipitous detection of \src in its quiescent state.
Two observations in 2012 (OsbIDs 00042728001 and 00042729001) covered the location of this source as part of the \textit{Swift} XRT Galactic plane survey program \citep{2013ATel.5200....1R}. 
The total exposure is 1055 seconds.
We find no X-ray source at this position within a 30\arcsec\ radius.
The 3$\sigma$ upper limit of the count rate is estimated to be $5.0\times 10^{-3}$ counts~sec$^{-1}$ within this radius. 
Then, 
the 3$\sigma$ upper limit of the 2--10~keV absorbed and absorption-corrected fluxes are calculated to be $5.0\times 10^{-13}$~erg~s~cm$^{-2}$ and $9.0\times 10^{-13}$~erg~s~cm$^{-2}$, respectively, on the assumption of an absorbed blackbody spectrum with  $N_{\rm H}=8.72\times 10^{22}$~cm$^{-2}$ and $kT=1.1$~keV, which correspond to an upper limit of the quiescent X-ray luminosity of $1\times 10^{34}$~erg~s$^{-1}$ at \distance.

The characteristic age of \src is inferred to be 2.04 kyr (see Table~1). If we assume that its true age is comparable to its characteristic age, we expect to find a young supernova remnant (SNR) surrounding the neutron star. 
Detection of an associated SNR, combined with the proper-motion measurement, would be helpful and can be important for constraining the magnetar's true age given that the characteristic age may be unreliable due to underlying assumptions about the neutron star rotation period at birth and braking index. 
Our search of archival radio, infrared, and X-ray data (e.g., \citealt{2019JApA...40...36G}) for a SNR or  pulsar-wind nebula coinciding at the position of \src, ($l$,$b$)=(327.872,$-0.335$) in the galactic coordinates using SkyView\footnote{https://skyview.gsfc.nasa.gov} fails to yield any convincing candidates. 

The celestial position of \src is close to another magnetar 1E~1547.0$-$5408, with an angular separation of about 0.7 degrees. The distance to 1E~1547.0$-$5408 is estimated to be 4--4.5~kpc from a dust scattering halo \citep{2010ApJ...710..227T} and a possible association with SNR~G327.24$-$0.13 \citep{2007ApJ...667.1111G}.
By contrast, no definitive information is available about the distance to \src. 
Taking into account  the fact that the column density of \src ($N_{\rm H}=8.7\times 10^{22}$~cm$^{-2}$) is larger than that of 1E~1547.0$-$5408 ($N_{\rm H}=3.2\times 10^{22}$~cm$^{-2}$; \citealt{2010PASJ...62..475E}), we assume a fiducial distance of \distance, which corresponds to the location on the Scutum–Centaurus Arm. 

Figure~\ref{fig:discussion}b compares pulsar X-ray luminosity $L_{\rm x}$ with  spin-down luminosity $L_{\rm sd}$. 
Immediately after the outburst, the surface emission (2--10\,keV) and total (2--60\,keV) luminosity of \src were $L_{\rm x}=8.5\times 10^{35}d_{10}^2$~erg~s$^{-1}$ and $9.6\times 10^{35}d_{10}^2$~erg~s$^{-1}$, respectively. 
At the assumed distance of \distance, these are respectively 40 and 46 times larger than its spin-down luminosity $L_{\rm sd}=2.1\times 10^{34}$~erg~s$^{-1}$. 
The observed X-ray luminosity is similar to those of past reported outbursts of transient magnetars. 
The upper limit on the quiescent X-ray luminosity of \src is $L_{\rm quie}=1\times 10^{34}d_{10}^2$, corresponding to $L_{\rm quie}/L_{\rm sd}<0.55$, which is  still up to 2$-$3 orders of magnitude higher than those of the rotation-powered pulsars 
(see Figure \ref{fig:discussion}b and Figure 12 of \citealt{2019RPPh...82j6901E}). 
The upper limit on the quiescent luminosity also makes \src one of the coolest young magnetars (see, e.g., \citealt{2015SSRv..191..239P}) and is compatible with $L\sim2\times 10^{33}$~erg~s$^{-1}$ of PSR~J1119$-$6127  \citep{2005ApJ...630..489G,2012ApJ...761...65N,2021arXiv210612018B}.  

\subsection{Peculiarities of the slow decline outburst} 
The persistent X-ray flux of most transient magnetars remains in a bright plateau state for a few weeks immediately after the onset of outbursts and then starts to fade over the next several months \citep{2018MNRAS.474..961C}. 
Typically, the plateau duration and decaying slope are $\tau_0=$11--43\,days and $p\sim$0.7--2 \citep{2017ApJS..231....8E}, respectively, when the X-ray flux $F_{\rm x}$ is fitted with an empirical formula $F_{\rm x}(t)=F_0/(1+t/\tau_0)^p$ where $t$, $F_0$ are the elapsed time and plateau flux, respectively.
A peculiarity of \src is its long-lasting outburst. 
The absorbed X-ray flux stayed nearly  stable at around $4\times 10^{-11}$\,erg~s$^{-1}$~cm$^{-2}$ with only a slow decline over a month. There was no apparent rollover of the flux trend as of \nicerlast. 
Such long-lasting outbursts are rare for magnetars with only a few exceptions ever recorded, e.g., the radio-loud magnetar 1E~1547.0$-$5408. 

Another peculiarity of the \src outburst is the higher temperature ($kT \sim$1.1\,keV) of the blackbody component than the typical value of known magnetars, $kT\sim 0.3$--0.7\,keV \citep{2017ApJS..231....8E}, despite  the surface-emission radius $R\sim 2d_{10}$~km of \src being well within the typical range for a neutron star.
During the one-month observation, the slow flux decline originated from the decreasing emission radius rather than the temperature decline.
This fact suggests a situation where the hot spot on the neutron star surface responsible for the X-ray emission was shrinking, which is consistent with the twisted magnetosphere model \citep{2017ARA&A..55..261K}. 

We detected 37 and 5 short bursts with \textit{NICER} and \textit{Swift}/BAT, respectively, during the initial two weeks since discovery of \src (Figure~\ref{fig:monitoring}j).
The burst-active periods of persistently bright magnetars (e.g., SGR~1806$-$20) are known to be longer ($\gtrsim$100 days) than those ($\lesssim$10 days) of low-burst rate transient ones (e.g., SGR~0501$+$4516) \citep{2014AN....335..296G}. 
The burst-active period of \src is close to the latter case.
We conjecture that repeated bursts as observed in \src would provide impulsive heating of the surface to sustain the long-lasting decay.
To investigate the potential relationship between the bursts and persistent emission on the pulse profile, we plot in Figure \ref{fig:pulse_profile}e the phase distribution of  the observed bursts. An Anderson Darling (AD) test suggests that the burst phase distribution differs from a uniform distribution with an AD statistic of 0.63 and corresponding p-value of 0.61. Thus there is neither a statistical difference between the phase distribution of the bursts and uniform distribution nor a statistically significant correlation between the burst occurrence and the pulse profile so far.

\subsection{Comparison with radio emitting magnetars \label{sec:compare_radio}}

We did not detect any radio emission from this source. 
It has been long established theoretically that as far as ordinary pulsars are concerned, the occurrence of rotation-powered polar-cap radio emission requires a  sufficiently large potential drop $\Delta\Phi$ to generate electron-positron pairs near polar caps \citep{1969ApJ...157..869G,1971ApJ...164..529S,1975ApJ...196...51R,1979ApJ...231..854A}; note that the conventional definition of $\Delta\Phi\propto L_{\rm sd}^{1/2}$ implies that radio emission is equivalently related to $L_{\rm sd}$ \citep{2013MNRAS.429..113H}. The observed radio luminosity of rotation-powered pulsars indeed follows the relation (e.g., \citealt{2002ApJ...568..289A}). In the absence of magnetically-induced non-potential fields, magnetars which satisfy this condition at their polar caps before crossing the death line should be in principle capable of producing coherent radio emission. 
However, as shown in Figure~\ref{fig:discussion}a-c and summarized in Appendix Table~\ref{tab:radio-loud_outburst}, pulsed radio emission has been only reported from 6 radio-loud magnetars and a high-B pulsar that exhibited a magnetar-like outburst.
It is an open question under what conditions a magnetar becomes radio-loud. The new magnetar \src is located in the $P$-$\dot{P}$ parameter space close to radio-loud magnetars (Figure~\ref{fig:discussion}a). The DSN upper limits on the radio flux (0.043~mJy for S-band and 0.026~mJy for X-band) would be much lower than the flux densities of the known radio-loud magnetars if located at the same distance of \src (10 kpc assumed), $\sim0.3-5$ mJy and $\sim0.1-30$ mJy at S-band and X-band, respectively \citep{2007ApJ...666L..93C,2007ApJ...669..561C,2008ApJ...679..681C,2010ApJ...721L..33L,2012MNRAS.422.2489L,2013MNRAS.435L..29S,2015ApJ...808...81P,2021MNRAS.505.1311H}.
Therefore, the lack of radio emission suggests the existence of some other  physical factors than simply $P$ and $\dot{P}$ that govern radio emission.

One crucial factor is the geometry among the pulsar rotation axis, magnetic axis, line of sight to the observer and the width of any putative radio beam. 
For example, an anisotropic radio beam aligned to the magnetic axis must cross the line of sight to be detected, and longer period rotation-powered pulsars tend to have narrower beams (e.g., \citealt{1988MNRAS.234..477L,1993ApJ...405..285R}). The radio non-detection of \src might suggest that this magnetar may not have a favorable geometry for detection (e.g., \citealt{2012ApJ...744...97L}). Some radio-loud magnetars have observational signatures that suggest aligned rotators (i.e., the angle between the magnetic axis and rotation axis is $\lesssim30^\circ$; \citealt{2008ApJ...679..681C,2012MNRAS.422.2489L,2020ApJ...896L..37L}). As shown in Figure~\ref{fig:discussion}c, the PF 
both in quiescence and during X-ray outbursts of radio-loud magnetars is lower than that of the other magnetars.
A low PF 
could be due to radio-loud magnetars being observed as near-aligned rotators.  
The same interpretation can be inferred from the systematic Fourier-decomposition study of X-ray profiles of magnetars in quiescence conducted by
\citet{2019MNRAS.485.4274H}; radio-loud magnetars have a more sinusoidal pulse profile with a pronounced first Fourier component accompanied with a weaker second component. 
Another factor for the radio emission is the effects of higher-order (and possibly not curl-free) magnetic field components other than the dipolar field inferred from $P$-$\dot{P}$, which has been already taken into account.
Quantum electrodynamic effects affect the conditions for pair cascades and radio emission when a magnetic field is sufficiently strong, above $10^{13}$ G. For example, if photon splitting occurs in a region above the critical magnetic field, the way the electron/positron cascade occurs is modified, it is perhaps quenched, and radio pulsation may be suppressed \citep[e.g.,][]{2001ApJ...547..929B}.
Finally, magnetar magnetospheres are considered to be dynamic and radio flux, pulse profile, and polarization swing pattern can change significantly within a few days (e.g., \citealt{2021MNRAS.502..127L}). 
Future monitoring of this new magnetar \src will be important for understanding the conditions for the magnetar radio emission.


\begin{acknowledgments}
The authors are grateful to the {\it NICER}, {\it NuSTAR}, {\it Swift}, and {\it DSN} scheduling and operation teams. We thank Francesco Coti Zelati, Alice Borghese, Nanda Rea, Gian Luca Israel, and Paolo Esposito for helpful discussion about the initial observation and for preparing the GCN reports. We thank Mareki Honma, Takaaki Jike, Aya Yamauchi, Toshio Terasawa, and Tomoya Hirota for useful comments on the VERA observation and all the staff of Mizusawa VLBI Observatory of NAOJ for operating the VERA array.
T.E. and S.K. are supported by JSPS/MEXT KAKENHI grant numbers 17K18776, 18H04584, 18H01246, 19K14712 and 21H01078.
A.B.P is a McGill Space Institute (MSI) Fellow and a Fonds de Recherche du Quebec -- Nature et Technologies (FRQNT) postdoctoral fellow.
W.C.G.H. acknowledges support through grant 80NSSC20K0278 from NASA. NICER research at NRL is supported by NASA.
W.A.M is grateful to the DSN scheduling team and the Canberra Deep Space Communication Complex~(CDSCC) staff for scheduling and carrying the radio observations with the DSN.
A portion of this research was performed at the Jet Propulsion Laboratory, California Institute of Technology, under a Research and Technology Development Grant through a contract with the National Aeronautics and Space Administration. U.S. government sponsorship is acknowledged.

\end{acknowledgments}



%

\vspace{5mm}
\facilities{{\it NICER}, {\it NuSTAR}, {\it 
Swift}, DSN, VERA}


\software{\texttt{HEASoft}, \texttt{astropy}, \texttt{PINT}}
          

\clearpage 

\begin{figure*}[h]
\begin{center}
\vspace{2mm}
\includegraphics[width=0.45\textwidth]{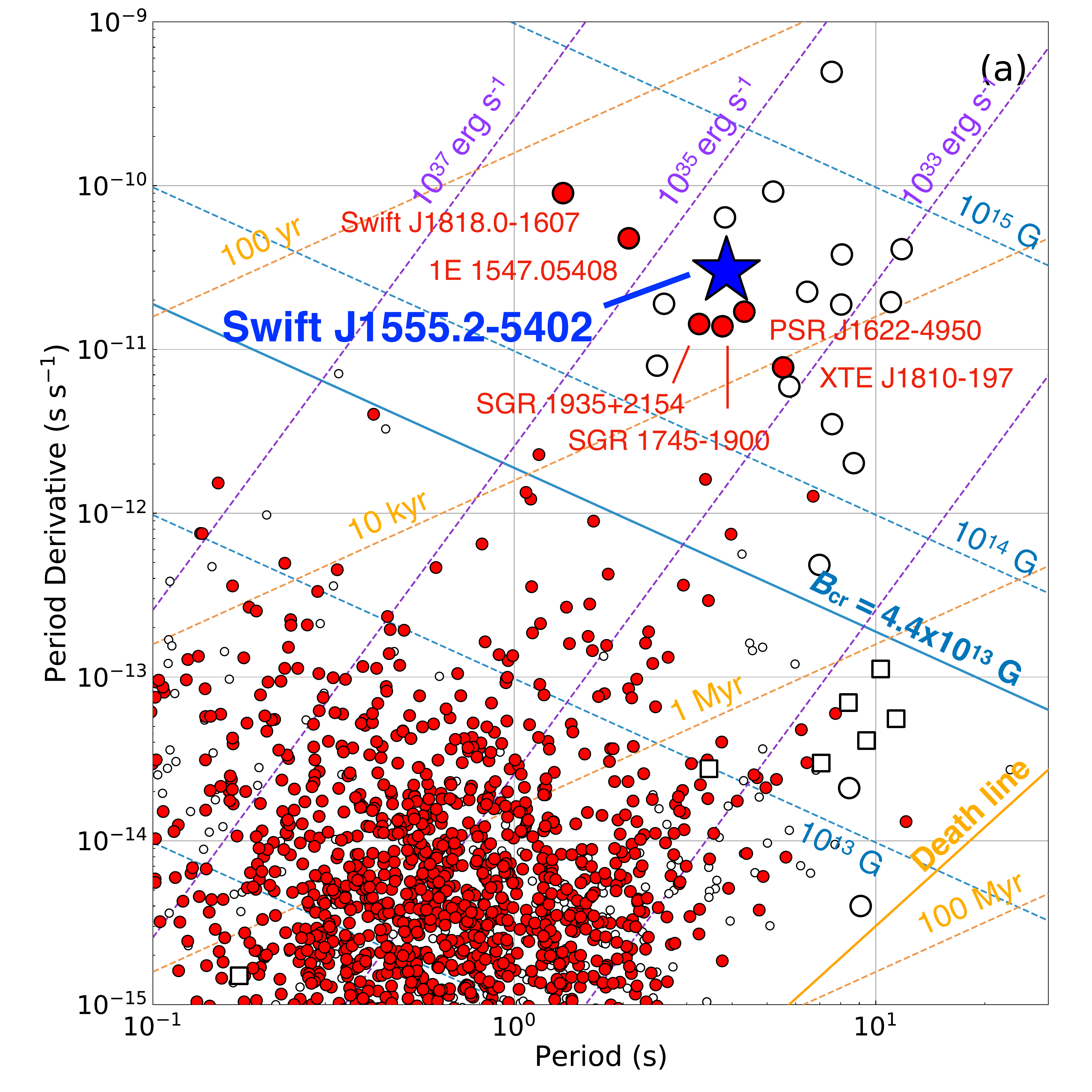}
\includegraphics[width=0.45\textwidth]{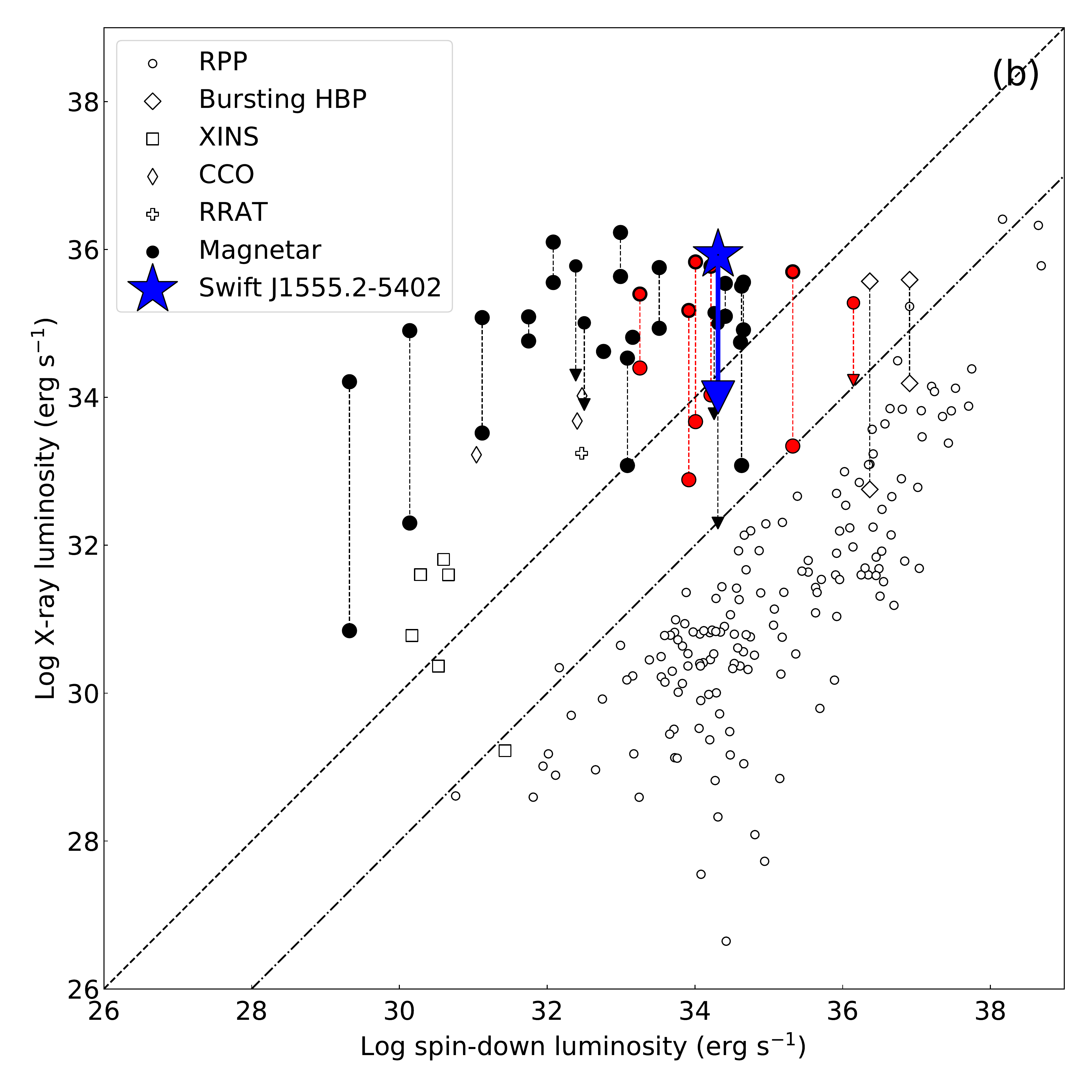}
\includegraphics[width=0.45\textwidth]{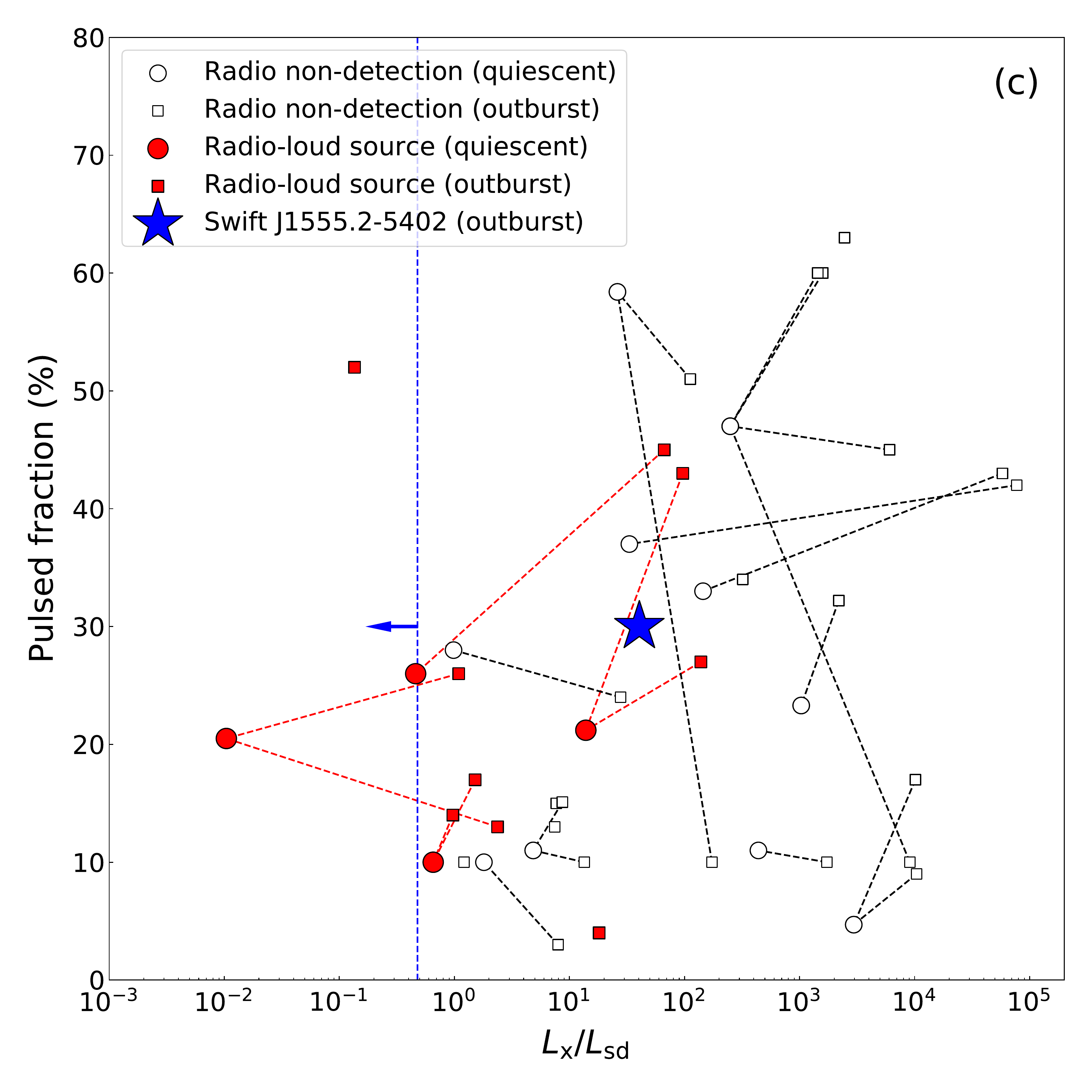}
\includegraphics[width=0.45\textwidth]{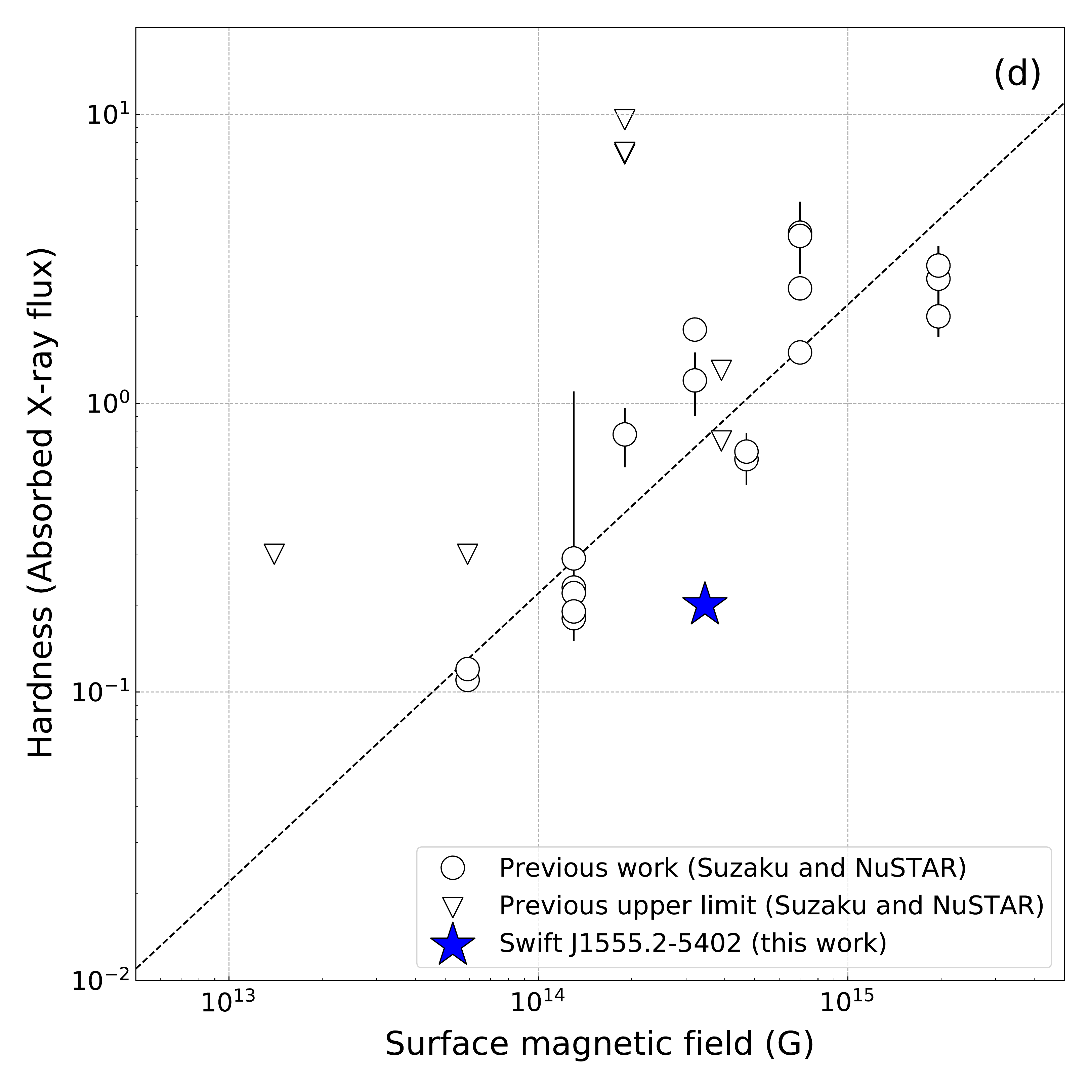}
\caption{
(a) Position of the new magnetar \src (star) on the $P$-$\dot{P}$ diagram. 
Large and small circles indicate magnetars (from the McGill catalog; \citealt{2014ApJS..212....6O}) and canonical rotation-powered pulsars (from the ATNF catalog; \citealt{2005AJ....129.1993M}), respectively.
Filled red symbols indicate radio-emitting pulsars. 
The lines show constant surface magnetic field strengths, characteristic ages, and spin-down luminosities. 
(b) Observed X-ray luminosity in the soft X-ray band (including the unpulsed component) compared with the spin-down power for various types of pulsars. 
The peak X-ray luminosity and quiescent values of magnetars are connected via dashed lines. Red symbols indicate radio-loud magnetars. 
The two diagonal lines indicate where the X-ray luminosity becomes equal to 100\% and 1\% of the spin-down power. 
The values and references used in panels (b)-(d) are summarized in Appendix Table~\ref{tab:radio-loud_outburst} and \ref{tab:radio-quiet_outburst} for magnetars and in \citet{2019RPPh...82j6901E} for other pulsars.
(c) Pulsed fractions as a function of the X-ray luminosity normalized by the spin-down power. Filled red symbols indicate radio-emitting magnetars. Circles and squares represent data in quiescence and during X-ray outbursts, respectively. A dashed line connects observations for the same source.
The vertical dashed line with the arrow indicates the region of the quiescent state of \src. 
(d) The broad-band hardness ratio of absorbed X-ray fluxes between the 1--10~keV and 15--60~keV with the best-fit correlation \citep{2017ApJS..231....8E} . 
\label{fig:discussion}}
\end{center}
\end{figure*}

\clearpage
\begin{deluxetable*}{lrrrr}
\tablecaption{Summary of timing and spectral properties of \src \label{tab:summary_parameter}}
\tablewidth{0pt}
\tablehead{
\colhead{Parameter} & \multicolumn{4}{c}{Values} 
}
\startdata
\multicolumn{5}{c}{Timing properties (NICER monitoring)} \\
\hline
MJD range & \multicolumn{4}{r}{59368.58--59396.98} \\ 
Epoch $T_0$ (MJD)  & \multicolumn{4}{r}{59382.7549} \\
Spin frequency $\nu$ (Hz) & \multicolumn{4}{r}{0.258997103(8)} \\
Frequency derivative $\dot{\nu}$ ($10^{-12}$ Hz~s$^{-1}$) & \multicolumn{4}{r}{-2.04(5)} \\
Second frequency derivative $\ddot{\nu}$ ($10^{-18}$ Hz~s$^{-2}$) & \multicolumn{4}{r}{-4.50(13)} \\
Third frequency derivative $\nu^{(3)}$ ($10^{-23}$ Hz~s$^{-3}$) & \multicolumn{4}{r}{-1.10(10)} \\
Fourth frequency derivative $\nu^{(4)}$ ($10^{-29}$ Hz~s$^{-4}$) & \multicolumn{4}{r}{3.59(15)} \\
Fifth frequency derivative $\nu^{(5)}$ ($10^{-34}$ Hz~s$^{-5}$) & \multicolumn{4}{r}{1.59(14)} \\
RMS residual (phase) & \multicolumn{4}{r}{0.014} \\
$\chi^2$/d.o.f. & \multicolumn{4}{r}{117.863/126} \\
\hline 
Period $P$ (sec) & \multicolumn{4}{r}{3.86104705(12)} \\
Period derivative $\dot{P}$ ($10^{-11}$~s~s$^{-1}$) & \multicolumn{4}{r}{3.05(7)} \\
Second period derivative $\ddot{P}$ ($10^{-17}$~s~s$^{-2}$) & \multicolumn{4}{r}{6.7(2)} \\
Third period derivative $P^{(3)}$ ($10^{-22}$~s~s$^{-3}$) & \multicolumn{4}{r}{1.63(15)} \\
Fourth period derivative $P^{(4)}$ ($10^{-28}$~s~s$^{-4}$) & \multicolumn{4}{r}{-5.3(2)} \\
Fifth period derivative $P^{(5)}$ ($10^{-33}$~s~s$^{-5}$) & \multicolumn{4}{r}{-2.4(2)} \\ 
Characteristic age $\tau_{\rm c}$ (kyr) & \multicolumn{4}{r}{2.01(5)} \\
Surface magnetic field $B_{\rm surf}$ ($10^{14}$~G) & \multicolumn{4}{r}{3.47(4)} \\ 
Spin-down luminosity $L_{\rm sd}$ ($10^{34}$~erg~s$^{-1}$) & \multicolumn{4}{r}{2.09(5)} \\
\hline \hline 
\multicolumn{5}{c}{Spectral properties (NICER+NuSTAR joint fit)} \\
\hline 
Joint observation numbers & 1 & 2 & 3 & Average \\ 
Observation date (MJD) & 59370 & 59374 & 59386 & --  \\ 
Column density $N_{\rm H}$ ($10^{22}$~cm$^{-2}$) & \multicolumn{3}{c}{8.72(8)} & 8.59(7) \\
Temperature $kT$ (keV) & 1.144(4) & 1.148(4) & 1.153(4) & 1.153(3) \\
Radius $R$ (km) & 2.10(2) & 2.05(2) & 2.04(2) & 2.06(3) \\
Photon index $\Gamma$ & 1.27(12) & 1.68(17) & 1.15(0.16) & 1.20(9) \\
Absorbed 2--10~keV flux ($10^{-12}$~erg~s~cm$^{-2}$) & 46.16$_{-0.38}^{+0.19}$ & 45.41$_{-0.55}^{+0.14}$ & 44.48$_{-0.45}^{+0.18}$ & 46.31$_{-0.26}^{+0.16}$ \\
Unabsorbed 2--10~keV flux ($10^{-12}$~erg~s$^{-1}$~cm$^{-2}$) & 70.60(32) & 69.39(35) & 67.80(36) & 70.0(3) \\
10--60~keV flux ($10^{-12}$~erg~s~cm$^{-2}$) & 9.32$_{-1.38}^{+0.14}$ & 7.73$_{-1.65}^{+0.08}$ & 8.72$_{-3.0}^{+0.10}$ & 9.04$_{-0.69}^{+0.14}$ \\
2--10~keV luminosity $L_{\rm x}$ ($10^{35}$~erg~s$^{-1}$) & 8.47 & 8.33 & 8.14 & 8.40 \\
10--60~keV luminosity $L_{\rm x}$ ($10^{35}$~erg~s$^{-1}$) & 1.11 & 0.93 & 1.05 & 1.09 \\
\hline
\multicolumn{5}{c}{Quiescent (Swift)} \\
\hline
Absorbed 2--10~keV flux ($10^{-12}$~erg~s~cm$^{-2}$) & \multicolumn{4}{c}{$<0.50$ (3$\sigma$)} \\
Unabsorbed 2--10~keV flux ($10^{-12}$~erg~s~cm$^{-2}$) & \multicolumn{4}{c}{$<0.90$ (3$\sigma$)} \\
2--10~keV luminosity $L_{\rm x}$ ($10^{35}$~erg~s$^{-1}$) & \multicolumn{4}{c}{$<0.1$ (3$\sigma$)} \\
\enddata
\tablecomments{\\
    The column density of the three joint \textit{NICER} and \textit{NuSTAR} spectral fitting is fixed to the same value.  
    X-ray luminosity and radius are calculated on an assumption of a fiducial distance of \distance. Quoted errors indicate the 68\% confidence limit. 
	}
\end{deluxetable*}

\clearpage
\appendix

\renewcommand{\thetable}{\Alph{section}\arabic{table}}
\renewcommand{\thefigure}{\Alph{section}\arabic{figure}}  

\setcounter{table}{0}
\setcounter{figure}{0}
\section{List of observations}
Tables \ref{tab_obsid_swift}, \ref{tab_obsid_nicer}, \ref{tab_obsid_nustar}, and \ref{tab_obsid_dsn} summarize the observations of \textit{Swift}, \textit{NICER}, \textit{NuSTAR}, and \textit{DSN}, respectively, conducted for this campaign as of \nicerlast.

\begin{deluxetable*}{ccccccrrr}
	\tabletypesize{\small}
	\tablecolumns{6}
	\tablewidth{0pt}
	\tablecaption{A list of \textit{Swift} ObsIDs of \src. \label{tab_obsid_swift}}
	\tablehead{
		\colhead{\#} & 
		\colhead{ObsID} & 
		\colhead{Mode} & 
		\colhead{Start Time} &
		\colhead{End Time} &
		\colhead{MJD} &		
		\colhead{Elapsed} &				
		\colhead{Exposure} &
		\colhead{Rate} 		
		\\
		\colhead{} & 
		\colhead{} & 
		\colhead{} & 
		\colhead{(UTC)} &
		\colhead{(UTC)} &
		\colhead{} & 
		\colhead{day} & 		
		\colhead{(sec)} &
		\colhead{(cps)} 		
	}
	\startdata
1 & 01053220000 & PC & 2021-06-03T09:46:24 & 2021-06-03T11:21:33 & 59368.463 & 0.06 & 1706 & $0.69\pm0.06$  \\
2 & 00014352001 & WT & 2021-06-04T7:50:02 & 2021-06-04T9:37:00 & 59369.364 & 1.0 & 1965 & $0.52\pm0.02$  \\
3 & 00014352002 & WT & 2021-06-05T10:40:43 & 2021-06-05T15:46:23 & 59370.551 & 2.1 & 4895 & $0.51\pm0.02$  \\
4 & 00014352003 & WT & 2021-06-07T12:06:00 & 2021-06-07T13:49:00 & 59372.540 & 4.1 & 1940 & $0.53\pm0.02$  \\
	\enddata
	\tablecomments{\\
		MJD: Middle of the start and end time of an observation.\\
		Elapsed day: Elapsed days from the first short burst at MJD 59368.40678 detected with {\it Swift}/BAT. \\ 
		Rate: Background subtracted 2--10~keV count rate of {\it Swift}.
	   }
\end{deluxetable*}
\begin{deluxetable*}{cccccrrr}
	\tabletypesize{\small}
	\tablecolumns{6}
	\tablewidth{0pt}
	\tablecaption{A list of \textit{NICER} ObsIDs of \src. \label{tab_obsid_nicer}}
	\tablehead{
		\colhead{\#} & 
		\colhead{ObsID} & 
		\colhead{Start Time} &
		\colhead{End Time} &
		\colhead{MJD} &		
		\colhead{Elapsed} &				
		\colhead{Exposure} &
		\colhead{Rate} 		
		\\
		\colhead{} & 
		\colhead{} & 
		\colhead{(UTC)} &
		\colhead{(UTC)} &
		\colhead{} & 
		\colhead{day} & 		
		\colhead{(sec)} &
		\colhead{(cps)} 		
	}
	\startdata
1 & 4202190101 & 2021-06-03T11:21:31 & 2021-06-03T18:52:20 & 59368.630 & 0.2 & 2367     &  3.74  \\
2 & 4560010101 & 2021-06-04T04:00:20 & 2021-06-04T14:54:24 & 59369.394 & 1.0 & 2201     &  3.91  \\
3 & 4560010102 & 2021-06-05T01:25:40 & 2021-06-05T23:32:00 & 59370.520 & 2.1 & 7235     &  3.96  \\
4 & 4560010103 & 2021-06-06T02:31:20 & 2021-06-06T17:31:40 & 59371.418 & 3.0 & 2068     &  3.52  \\
5 & 4560010104 & 2021-06-07T02:42:51 & 2021-06-07T23:00:40 & 59372.536 & 4.1 & 2199     &  4.07  \\
6 & 4560010105 & 2021-06-08T00:24:27 & 2021-06-08T22:16:20 & 59373.472 & 5.1 & 2931     &  3.88  \\
7 & 4560010201 & 2021-06-09T04:16:47 & 2021-06-09T16:53:20 & 59374.441 & 6.0 & 6070     &  3.9  \\
8 & 4560010202 & 2021-06-10T01:56:20 & 2021-06-10T14:36:20 & 59375.345 & 6.9 & 2582     &  3.94  \\
9 & 4560010301 & 2021-06-11T02:43:40 & 2021-06-11T23:10:00 & 59376.539 & 8.1 & 1852     &  4.28  \\
10 & 4560010601 & 2021-06-12T08:11:19 & 2021-06-12T08:27:40 & 59377.347 & 8.9 & 910     &  4.18  \\
11 & 4560010401 & 2021-06-13T05:52:25 & 2021-06-13T21:38:20 & 59378.573 & 10.2 & 988    &  3.82  \\
12 & 4560010402 & 2021-06-14T08:12:26 & 2021-06-14T20:52:40 & 59379.606 & 11.2 & 915    &  3.99  \\
13 & 4560010501 & 2021-06-15T15:11:28 & 2021-06-15T15:28:00 & 59380.639 & 12.2 & 909    &  4.53  \\
14 & 4560010502 & 2021-06-16T06:41:09 & 2021-06-16T08:24:28 & 59381.314 & 12.9 & 1085   &  4.25  \\
15 & 4560010602 & 2021-06-17T02:49:31 & 2021-06-17T20:08:40 & 59382.479 & 14.1 & 911    &  3.3  \\
16 & 4560010603 & 2021-06-18T08:15:34 & 2021-06-18T19:22:40 & 59383.576 & 15.2 & 998    &  3.82  \\
17 & 4560010701 & 2021-06-19T04:24:22 & 2021-06-19T06:09:08 & 59384.220 & 15.8 & 1575   &  3.7  \\
18 & 4560010702 & 2021-06-20T05:10:51 & 2021-06-20T06:54:08 & 59385.252 & 16.8 & 1472   &  3.77  \\
19 & 4560010801 & 2021-06-21T05:58:08 & 2021-06-21T23:17:20 & 59386.610 & 18.2 & 3756   &  3.66  \\
20 & 4560010802 & 2021-06-22T00:33:27 & 2021-06-22T10:08:00 & 59387.223 & 18.8 & 4545   &  3.64  \\
21 & 4560010901 & 2021-06-23T05:59:25 & 2021-06-23T09:17:56 & 59388.319 & 19.9 & 1979   &  3.71  \\
22 & 4560010902 & 2021-06-24T03:41:05 & 2021-06-24T07:01:43 & 59389.223 & 20.8 & 2307   &  3.68  \\
23 & 4560011001 & 2021-06-26T00:34:00 & 2021-06-26T00:53:20 & 59391.030 & 22.6 & 902    &  3.36  \\
24 & 4560011002 & 2021-06-27T06:05:15 & 2021-06-27T06:19:40 & 59392.259 & 23.9 & 663    &  3.49  \\
25 & 4560011101 & 2021-06-27T07:38:15 & 2021-06-27T07:44:28 & 59392.320 & 23.9 & 176    &  3.3  \\
26 & 4560011102 & 2021-06-28T00:41:00 & 2021-06-28T02:28:20 & 59393.066 & 24.7 & 1271   &  3.21  \\
27 & 4560011201 & 2021-06-29T20:04:42 & 2021-06-29T23:20:24 & 59394.905 & 26.5 & 367    &  3.12  \\
28 & 4560011301 & 2021-07-01T21:35:48 & 2021-07-01T23:25:19 & 59396.938 & 28.5 & 1237   &  3.31  \\
	\enddata
	\tablecomments{\\
		MJD: Middle of the start and end time of an observation.\\
		Elapsed day: Elapsed days from the first short burst at MJD 59368.40678 detected with {\it Swift}/BAT. \\ 
		Rate: Background subtracted 2--10~keV count rate of {\it NICER}.
	   }
\end{deluxetable*}
\begin{deluxetable*}{lcccccrr}
	\tabletypesize{\small}
	\tablecolumns{6}
	\tablewidth{0pt}
	\tablecaption{A list of \textit{NuSTAR} ObsIDs of \src. \label{tab_obsid_nustar}}
	\tablehead{
	    \colhead{\#} &
		\colhead{ObsID} & 
		\colhead{Start Time} &
		\colhead{End Time} &
		\colhead{MJD start} &		
		\colhead{Exposure} & 
		\colhead{Rate A} & 
		\colhead{Rate B}  		
		\\
		\colhead{} & 
		\colhead{} & 
		\colhead{(UTC)} &
		\colhead{(UTC)} &
		\colhead{} & 
		\colhead{(ks)} & 	
		\colhead{(cps)} & 
		\colhead{(cps)}  		
	}
\startdata
1 & 90701319002 & 2021-06-05T10:20:48 & 2021-06-06T06:33:07 &59370.43111111 & 38.4 & 1.12& 1.03\\ 
2 & 80702313002 & 2021-06-09T05:39:00 & 2021-06-09T18:00:00 & 59374.23541667
& 25.0 & 1.09& 1.025\\
3 & 80702313004 & 2021-06-21T14:05:11 & 2021-06-22T04:15:00 &59386.58693287 & 28.9 & 0.99& 0.94 
\enddata
	\tablecomments{
		Rate A and B are the 3--79~keV count rate of \textit{NuSTAR} FPMA and FPMB, respectively. 
		}
\end{deluxetable*}
\begin{deluxetable}{cccccccc}
	\tabletypesize{\small}
	\tablecolumns{8}
	\tablewidth{0pt}
	\tablecaption{A list of radio observations of \src\ with the Deep Space Network~(DSN). \label{tab_obsid_dsn}}
	\tablehead{
		\colhead{\#} & 
		\colhead{Instrument} & 
		\colhead{Observation} &
		\colhead{Observation} &
		\colhead{Duration} & 
		\colhead{Observing} &
		\colhead{Mean Flux Density$^{\mathrm{b}}$} &
		\colhead{Radio Burst Fluence$^{\mathrm{c}}$} \\ [-0.25cm]
		\colhead{} & 
		\colhead{} & 
		\colhead{Start Time} &
		\colhead{Start Time} &
		\colhead{} & 
		\colhead{Frequency Band$^{\mathrm{a}}$} &
		\colhead{($S_{\text{mean}}^{\text{S-band}}$ / $S_{\text{mean}}^{\text{X-band}}$)} &
		\colhead{($\mathcal{F}^{\text{S-band}}$ / $\mathcal{F}^{\text{X-band}}$)} \\ [-0.25cm]
		\colhead{} &
		\colhead{} &
		\colhead{(UTC)} & 
		\colhead{(MJD)} & 
		\colhead{(Hours)} &
		\colhead{} &
		\colhead{(mJy / mJy)} &
		\colhead{(Jy ms / Jy ms)}
	}
	\startdata
	1 & DSN (DSS-36) & 2021 June 04 06:59:02 & 59369.29099 & 1.3 & S-band / X-band & $<$ 0.27 / $<$ 0.10   & $<$ 1.7 / $<$ 0.7 \\
	2 & DSN (DSS-36) & 2021 June 05 11:46:10 & 59370.49039 & 2.4 & S-band / X-band & $<$ 0.18 / $<$ 0.07   & $<$ 1.6 / $<$ 0.61 \\
	3 & DSN (DSS-43) & 2021 June 06 04:49:06 & 59371.20076 & 2.2 & S-band / X-band & $<$ 0.042 / $<$ 0.025 & $<$ 0.35 / $<$ 0.21 \\
	4 & DSN (DSS-34) & 2021 June 10 11:31:12 & 59375.48000 & 2.4 & S-band / X-band & $<$ 0.20 / $<$ 0.11   & $<$ 1.7 / $<$ 0.9 \\
	5 & DSN (DSS-43) & 2021 June 12 04:08:52 & 59377.17282 & 2.5 & S-band / X-band & $<$ 0.043 / $<$ 0.026 & $<$ 0.39 / $<$ 0.24 \\
	\enddata
	\tablecomments{\\
	    The first two observations (2021 June 4 and 5) and the fourth observation (2021 June 10) were carried out using DSS-36 and DSS-34, two 34 m diameter radio telescopes in Canberra, Australia, whereas the remaining observations (2021 June 6, 12, and 16) were carried out using DSS-43, the 70 m diameter dish in Canberra. 
		$^{\mathrm{a}}$ The center frequencies at S/X-band are 2.2/8.4\,GHz, respectively. \\
		$^{\mathrm{b}}$ 7$\sigma$ upper limits on the mean flux density in each radio frequency band, assuming a 10\% duty cycle. The uncertainties on the mean flux density upper limits are estimated at 15\%, primarily due to the uncertainty in the system temperature. \\
		$^{\mathrm{c}}$ 7$\sigma$ upper limits on the radio burst fluence in each radio frequency band, assuming a burst width of 1\,ms. The uncertainties on the fluence detection thresholds are estimated at 15\%, primarily due to the uncertainty in the system temperature.}
\end{deluxetable}

\clearpage
\setcounter{table}{0}
\setcounter{figure}{0}
\section{Burst analyses}
\begin{figure*}[h]
\begin{center}
\vspace{2mm}
\includegraphics[width=0.48\textwidth]{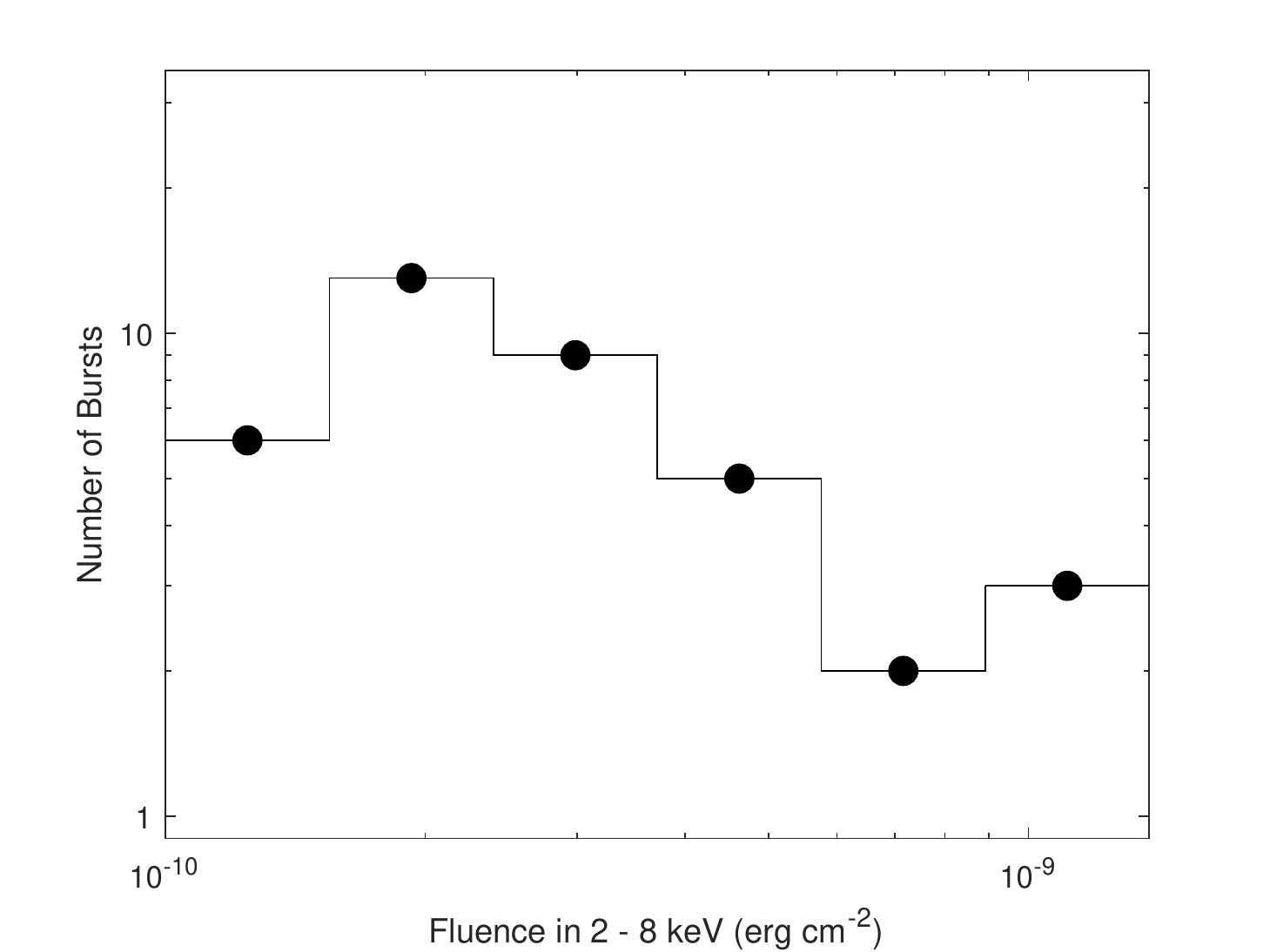}
\caption{
Fluence distribution in the 2--8\,keV of the detected short bursts from \src.
\label{fig:fluence}}
\vspace{-0.4cm}
 \end{center}
\end{figure*}

Tables \ref{tab_burst_bat} and \ref{tab_burst_nicer} summarize the detected magnetar short bursts by \textit{Swift}/BAT and \textit{NICER}, respectively. The corresponding fluence distribution is shown in Figure~\ref{fig:fluence}. 

\begin{figure*}[h]
\begin{center}
\vspace{2mm}
\includegraphics[width=0.48\textwidth]{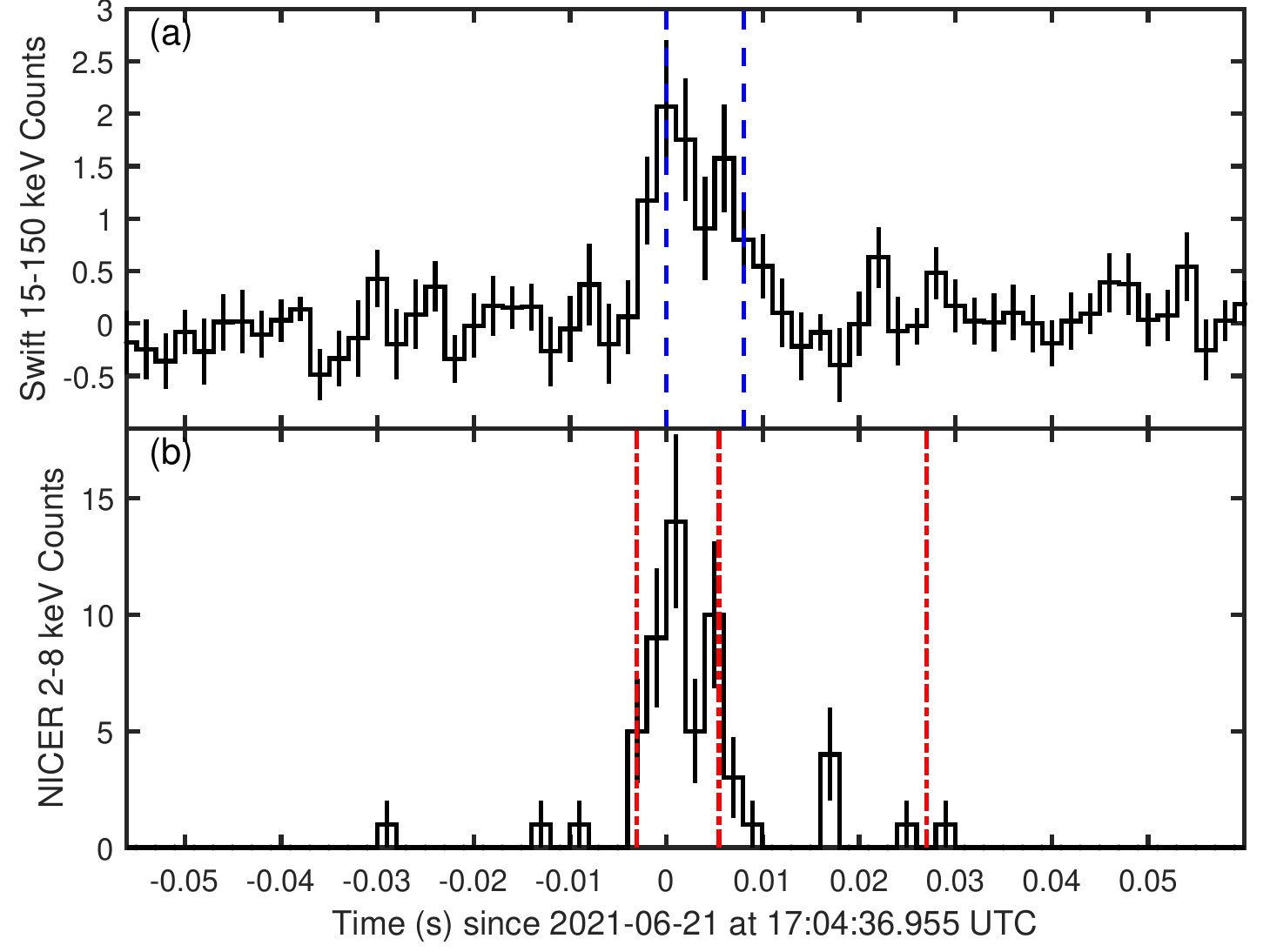}
\includegraphics[width=0.48\textwidth]{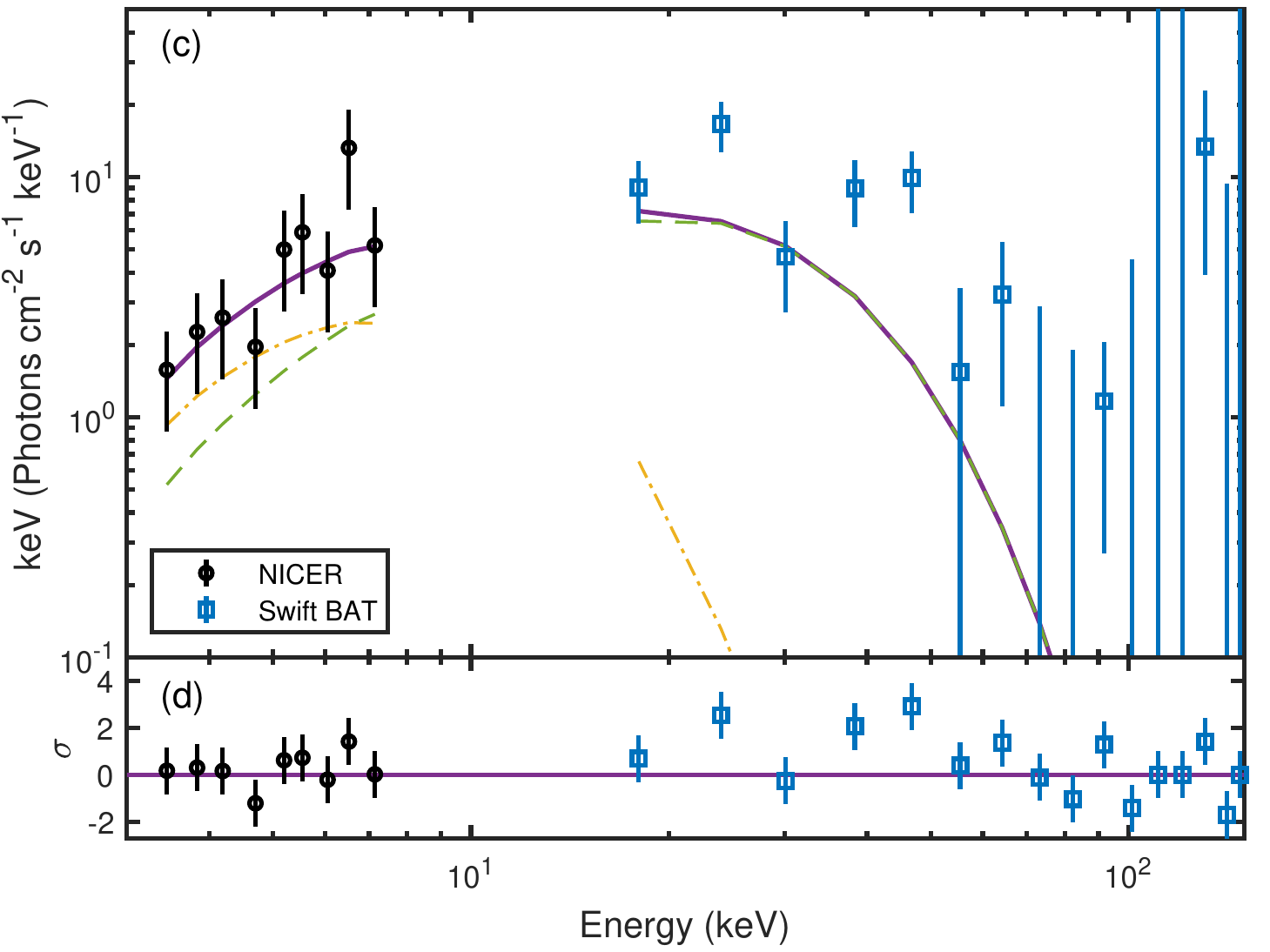}
\caption{
Light curves of a burst detected simultaneously with (a) \emph{Swift} BAT in 15--150 keV and (b) \emph{NICER} in 2--8 keV. The bin size of the light curves are 2~ms. Blue dashed lines denotes the start and the end of $T_{90}$ of the \swift\ light curve, where the red dashed-dotted lines are the boundaries of Bayesian blocks detected with \emph{NICER} events. Broadband X-ray spectrum of this burst is shown in panel (c) with the best-fit model (purple solid line) consisting of two blockbodies (orange dashed-dotted line and green dashed line). The residual is shown in panel (d). 
\label{fig:BAT_bursts}}
\vspace{-0.4cm}
 \end{center}
\end{figure*}

A burst (the burst number 5 in Table \ref{tab_burst_bat} and 26 in Table \ref{tab_burst_nicer}) is simultaneously detected with \swift\ BAT and \emph{NICER}. The light curves of this simultaneous event are shown in Figure \ref{fig:BAT_bursts} (a) and (b). We extracted the broadband X-ray spectrum and fit it with an absorbed power law and set $N_{\textrm{H}}=8.72\times10^{22}$ cm$^{-2}$ (See Figure \ref{fig:BAT_bursts}c and d). The spectrum can be fitted by two blackbodies although the normalization cannot be well constrained. The soft one has the temperature of $kT=2.5_{-0.7}^{+3.3}$ keV, which is set to be 0.37 times the temperature of the hard component \citep{2009PASJ...61..109N}.    On the other hand, the spectrum can also be well fitted by an absorbed blackbody with $kT=5.3\pm0.8$ keV and a radius of $7.0\pm1.6$~k at \distance.

The fluence distribution is power-law like, but the index cannot be well constrained due to limited sample (Figure~\ref{fig:fluence}). We applies the Anderson Darling (AD) test to assess the observed fluence distribution against a power-law distribution with a index of $-1$. This yields an AD statistic of 2.3 with a corresponding p-value of 0.04. In comparison, we perform the same test against a uniform distribution and obtain an AD statistic of 78 with a p-value of $1\times10^{-5}$. This suggest that the fluence is not uniformly distributed.

\begin{deluxetable*}{cccccccc}
	\tabletypesize{\small}
	\tablecolumns{6}
	\tablewidth{0pt}
	\tablecaption{A list of short bursts from \src ~detected with \textit{Swift}/BAT. \label{tab_burst_bat}}
	\tablehead{
		\colhead{\#} & 
		\colhead{Trigger ID} & 
		\colhead{Time} &
		\colhead{Duration} &
		\colhead{SNR} &		
		\colhead{$kT$} & 
		\colhead{fluence} & 
		\colhead{$\chi^2$} 
		\\
		\colhead{} & 
		\colhead{} & 
		\colhead{(UTC)} &
		\colhead{$T_{90}$ (ms)} &
		\colhead{} & 
		\colhead{(keV)} &
		\colhead{} & 
		\colhead{} 
	}
	\startdata
1 & 1053220 & 2021-06-03T09:45:46.589 & $12\pm 2.8$ & 9.9 & $6.66\pm 0.98$ & $9.09\pm 2.32$ & $33.98$ \\	
2 & 1053653 & 2021-06-05T23:52:04.582 & $14\pm 4.5$ & 7.3 & $8.53\pm1.40$ & $7.47\pm 2.62$ & 28.54  \\
3 & 1053961 & 2021-06-07T12:33:40.020 & $4\pm 2.2$ & 5.0 & N/A$^{\star}$ & N/A$^{\star}$ & N/A$^{\star}$ \\
4 & 1056025 & 2021-06-16T14:44:30.489 & $7\pm 2.8$ & 6.9 & $6.58\pm 2.22$ & $<3.08$  & 31.20 \\
5 & 1057131 & 2021-06-21T17:04:36.839 & $12\pm 4.5$ & 6.9 & $6.47\pm 2.03$ & $<3.82$ & 55.54 \\ 
	\enddata
	\tablecomments{\\
	Reported errors are 90\% confidence for each parameter.\\
	Time: Burst detection time (UTC) determined as the start time of $T_{90}$. \\
	SNR: Signal-to-noise ratio (SNR) of the BAT image in the 15--350~keV. \\
	kT: Blackbody temperature (keV) when fitted by the single blackbody model. \\ 
	fluence: Burst fluence in the 15--150~keV band ($10^{-9}$~erg~cm$^{-2}$). \\
	$\chi^2$: fitting chi-square values for 57 degree of freedom. \\
	$^{\star}$Burst \#3 is too weak to constrain spectral-fit parameters. \\
	}
\end{deluxetable*}

\begin{deluxetable*}{ccccccc}
	\tabletypesize{\small}
	\tablecolumns{6}
	\tablewidth{0pt}
	\tablecaption{A list of short bursts from \src ~detected with \textit{NICER}. \label{tab_burst_nicer}}
	\tablehead{
		\colhead{\#} & 
		\colhead{ObsID} & 
		\colhead{Time} &
		\colhead{Duration} &
		\colhead{Significance} &		
		\colhead{Phase} & 
		\colhead{Fluence}
		\\
		\colhead{} & 
		\colhead{} & 
		\colhead{(UTC)} &
		\colhead{(ms)} &
		\colhead{$\sigma$} & 
		\colhead{} &
		\colhead{}
	}
	\startdata
1 & 4202190101 & 2021-06-03T13:51:09.341 & 14.94 & 8.80 & 0.085 & $2.6\pm0.8$\\ 
2 & 4202190101 & 2021-06-03T13:51:27.181 & 59.18 & 6.38 & 0.711 & $2.4\pm0.8$\\ 
3 & 4202190101 & 2021-06-03T13:54:40.275 & 19.79 & 15.47 & 0.717 & $6.3\pm1.3$\\ 
4 & 4202190101 & 2021-06-03T18:49:06.251 & 35.90 & 6.35 & 0.166 & $1.9\pm0.7$\\ 
5 & 4560010101 & 2021-06-04T14:37:30.785 & 18.43 & 6.91 & 0.874 & $1.9\pm0.7$\\ 
6 & 4560010101 & 2021-06-04T14:43:50.878 & 15.12 & 11.32 & 0.316 & $4.0\pm1.0$\\ 
7 & 4560010102 & 2021-06-05T05:57:03.457 & 49.87 & 6.19 & 0.449 & $2.1\pm0.7$\\ 
8$^{\star}$ & 4560010102 & 2021-06-05T13:55:42.156& 15.91 & 5.67 & 0.504 & $1.3\pm0.6$\\ 
9 & 4560010102 & 2021-06-05T23:31:46.102 & 34.26 & 6.50 & 0.454 & $2.4\pm0.8$\\ 
10 & 4560010105 & 2021-06-08T11:19:42.895 & 9.55 & 7.63 & 0.285 & $1.9\pm0.7$\\ 
11 & 4560010105 & 2021-06-08T11:23:17.712 & 3.20 & 5.95 & 0.922 & $1.0\pm0.5$\\ 
12 & 4560010201 & 2021-06-09T05:51:19.191 & 19.64 & 10.85 & 0.299 & $4.2\pm1.1$\\ 
13 & 4560010201 & 2021-06-09T07:24:11.030 & 7.34 & 6.13 & 0.375 & $1.3\pm0.6$\\ 
14 & 4560010201 & 2021-06-09T10:34:21.954 & 7.21 & 6.50 & 0.747 & $1.3\pm0.6$\\ 
15 & 4560010201 & 2021-06-09T13:41:17.195 & 25.52 & 8.77 & 0.438 & $3.2\pm0.9$\\ 
16 & 4560010201 & 2021-06-09T13:42:48.797 & 19.87 & 6.18 & 0.162 & $1.6\pm0.6$\\ 
17 & 4560010202 & 2021-06-10T02:00:03.424 & 14.08 & 6.89 & 0.687 & $1.9\pm0.7$\\ 
18 & 4560010202 & 2021-06-10T05:05:18.924 & 21.26 & 6.24 & 0.540 & $1.9\pm0.7$\\ 
19 & 4560010301 & 2021-06-11T10:39:52.520 & 6.94 & 6.11 & 0.570 & $1.3\pm0.6$\\ 
20 & 4560010602 & 2021-06-17T02:51:59.547 & 6.71 & 16.58 & 0.542 & $6.6\pm1.3$\\ 
21 & 4560010602 & 2021-06-17T02:53:19.142 & 54.05 & 19.37 & 0.162 & $13.2\pm1.9$\\ 
22 & 4560010602 & 2021-06-17T02:59:19.487 & 72.30 & 7.00 & 0.493 & $2.6\pm0.8$\\ 
23 & 4560010701 & 2021-06-19T04:33:05.105 & 7.04 & 7.19 & 0.028 & $1.6\pm0.6$\\ 
24 & 4560010701 & 2021-06-19T04:40:15.346 & 7.71 & 7.19 & 0.457 & $1.9\pm0.7$\\ 
25 & 4560010701 & 2021-06-19T06:07:18.896 & 23.03 & 9.82 & 0.305 & $3.4\pm1.0$\\ 
26$^{\star\star}$ & 4560010801 & 2021-06-21T17:04:36.952 & 8.54 & 20.58 & 0.664 & $10.8\pm1.7$\\ 
26-1$^{\dagger}$ & 4560010801 & 2021-06-21T17:04:36.960 & 21.55 & 6.25 & 0.668 & $2.6\pm0.8$\\ 
27 & 4560010801 & 2021-06-21T23:05:26.979 & 17.84 & 6.39 & 0.767 & $1.6\pm0.6$\\ 
28 & 4560010801 & 2021-06-21T23:06:53.759 & 35.45 & 8.78 & 0.244 & $3.2\pm0.9$\\ 
29 & 4560010802 & 2021-06-22T00:35:18.974 & 12.50 & 8.94 & 0.228 & $2.4\pm0.8$\\ 
30 & 4560010802 & 2021-06-22T00:45:29.330 & 15.47 & 8.55 & 0.307 & $2.6\pm0.8$\\ 
31 & 4560010802 & 2021-06-22T02:20:44.171 & 32.71 & 8.16 & 0.385 & $2.9\pm0.9$\\ 
32 & 4560010802 & 2021-06-22T03:55:16.474 & 30.36 & 10.37 & 0.445 & $4.0\pm1.0$\\ 
33 & 4560010802 & 2021-06-22T08:26:59.567 & 6.15 & 12.71 & 0.746 & $4.0\pm1.0$\\ 
34 & 4560010802 & 2021-06-22T08:29:34.911 & 49.79 & 19.15 & 0.984 & $10.6\pm1.7$\\ 
35 & 4560010802 & 2021-06-22T10:00:07.220 & 15.05 & 8.79 & 0.882 & $2.6\pm0.8$\\ 
36 & 4560010902 & 2021-06-24T06:50:27.424 & 5.97 & 5.68 & 0.654 & $1.1\pm0.5$\\ 
37 & 4560011001 & 2021-06-26T00:52:34.199 & 12.48 & 9.24 & 0.167 & $2.6\pm0.8$\\
38 & 4560011301 & 2021-07-01 23:21:28.122 & 5.28 & 12.89 & 0.860 & $4.2\pm1.1$ 
	\enddata
	\tablecomments{\\
	Time: Burst detection time (UTC) determined as the start time of the Bayesian block. \\
	Duration: duration between two consecutive Bayesian blocks. \\
	Significance: Detection significance of the burst from Poisson-distributed noise. \\
	Fluence: Burst fluence in the 2--8~keV band ($10^{-10}$~erg~cm$^{-2}$) estimated from the number of photons and assuming a blackbody spectrum with $kT=2.8$~keV. The uncertainty is simply calculated from the Poisson noise, i.e., the square root of the number of photons.\\
	$^{\star}$ This burst was simultaneously observed with the second observation of DSN (Table \ref{tab_obsid_dsn}).\\
	$^{\star\star}$ This burst is simultaneously observed with {\it Swift} BAT (burst \#5 in Table \ref{tab_burst_bat}).\\
	$^{\dagger}$ This candidate is the tail of burst \#26.  \\
	}
\end{deluxetable*}

\clearpage
\setcounter{table}{0}
\setcounter{figure}{0}
\section{Glitch model} 
\label{sec:appendixtiming}

\begin{figure*}[h]
\begin{center}
\vspace{2mm}
\includegraphics[width=0.8\textwidth]{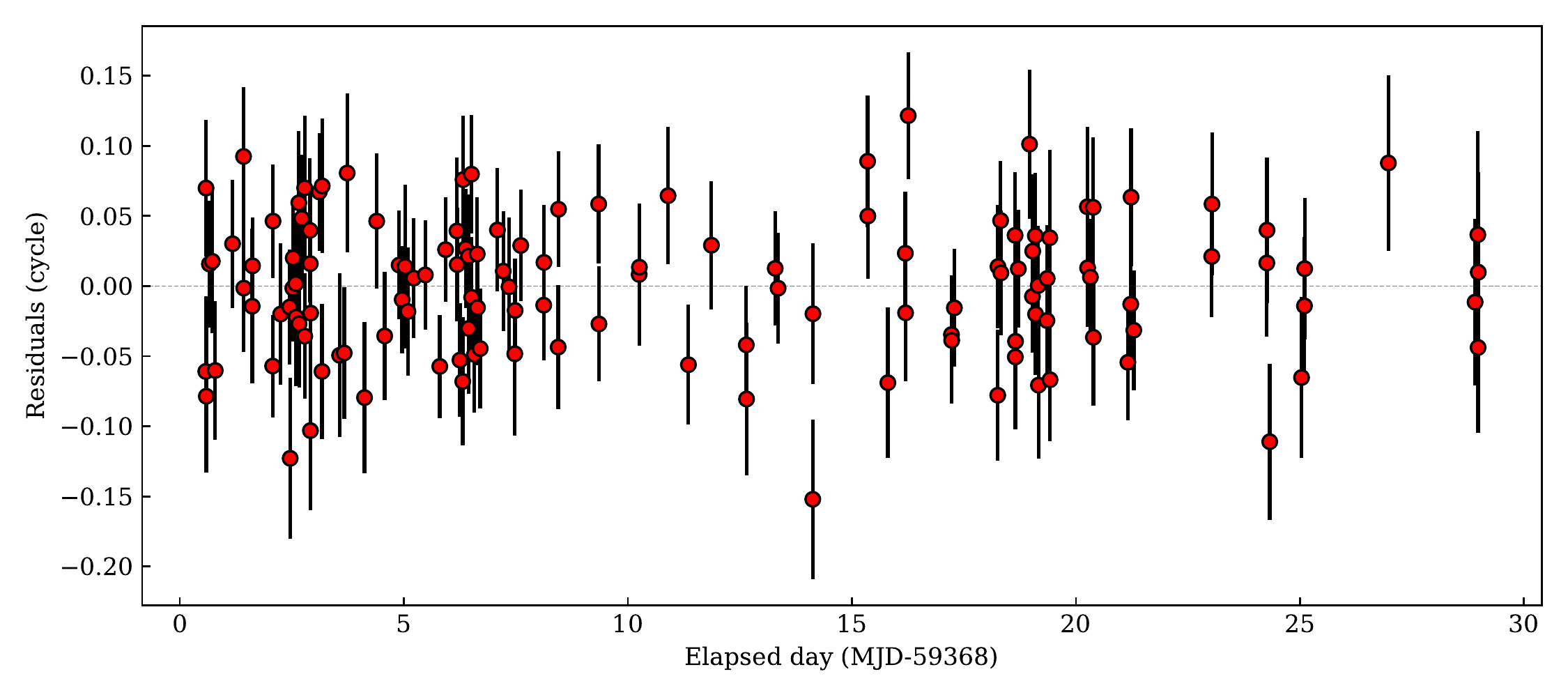}
\caption{Phase residuals after correcting for the glitch model presented in Table \ref{tab:summary_glitch}}
\label{fig:glitch}
\vspace{-0.4cm}
 \end{center}
\end{figure*}

We find that the timing behavior of \src is also well-described by a glitch model with three glitch candidates at MJDs 59373.6412, 59382.9236, and 59390.1618. The best-fit timing parameters are $\nu = 0.25899725(18)$ Hz and $\dot{\nu} = -1.65(18)\times10^{-12}$ Hz s$^{-1}$ at barycentric epoch $T_0 = {\rm MJD}\ 59382.7459$; the timing parameters and the glitch parameters are summarized in Table \ref{tab:summary_glitch}. The pulse phase residuals are characterized by an rms residual of $\sim0.012$ cycles, and they are plotted in Figure \ref{fig:glitch}. While the data can be plausibly described with a glitch model, it is more likely that the source is exhibiting strong timing noise (see Section \ref{sec:Discussion}), similar to that of Swift J1818.0-1607 \citep{2020ApJ...902....1H}.

The glitch sizes exhibited by \src are well within the range of observed values in magnetars, whereas the $\Delta\dot{\nu}$ values are among the highest values relative to observed values in magnetars \citep{2019AN....340..340H}. However, we are observing \src in outburst, and the high $\Delta\dot{\nu}$ values associated with the three glitch candidates are similar to that observed in some magnetars in outbursts as well \citep{2019AN....340..340H}. We also note the short recurrence timescale of the glitches (on the order of $\sim 10$ days). 

The timing features of \src show some similarities to that of Swift J1818.0$-$1607 - the two ``timing anomalies", were characterized as candidate glitches separated by 6 days, a traditional spin-up glitch of size $\Delta \nu = 2.7\times10^{-6}$ Hz and an anti-glitch with $\Delta \nu = -5.28\times10^{-6}$ Hz. We also observe a similar and unusual ``sign-switching" behavior of the glitches for \src, with an anti-glitch, a glitch, and another anti-glitch. 

\begin{deluxetable*}{lrrr}
\tablecaption{Summary of the glitch model with 3 glitch candidates for \src \label{tab:summary_glitch}}
\tablewidth{0pt}
\tablehead{
\colhead{Parameter} & \multicolumn{3}{c}{Values} 
}
\startdata
MJD range & \multicolumn{3}{c}{59368.53--59396.98} \\ 
Epoch $T_0$ (MJD) & \multicolumn{3}{c}{59382.7549} \\
Spin frequency $\nu$ (Hz) & \multicolumn{3}{c}{0.25899725(18)}  \\
Frequency derivative $\dot{\nu}$ (Hz s$^{-1}$) &  \multicolumn{3}{c}{$-1.65(18)\times10^{-12}$} \\
RMS residual (phase) & \multicolumn{3}{c}{0.012} \\
$\chi^2$/d.o.f. & \multicolumn{3}{c}{133.196/121} \\
\hline \hline
Glitch Candidate & 1 & 2 & 3 \\ 
\hline 
Glitch Epoch (MJD) & 59373.6412 & 59382.9236 & 59390.1618 \\
$\Delta\nu$ (Hz) & $-3.2(6)\times10^{-7}$ & $5(6)\times10^{-8}$ & $-6.9(1.0)\times10^{-7}$ \\
$\Delta\dot{\nu}$ (Hz~s$^{-1}$) & $3.3(1.7)\times10^{-13}$ & $-2.20(12)\times10^{-12}$ & $2.89(17)\times10^{-12}$ \\
\hline 
$\Delta\nu/\nu$ & $-1.2(0.2)\times10^{-6}$ & $2.1(2.5)\times10^{-7}$ & $-2.7(4)\times10^{-6}$ \\
$\Delta\dot{\nu}/\dot{\nu}$ & $-0.20(11)$ & 1.33(16) & $-1.7(2)$ \\
\enddata 
\end{deluxetable*}


\clearpage
\setcounter{table}{0}
\setcounter{figure}{0}
\section{Comparison with previous magnetar outbursts}

Tables \ref{tab:radio-loud_outburst} and \ref{tab:radio-quiet_outburst} summarize properties of previous magnetar outbursts. 

\begin{deluxetable*}{llrrrrrr}
	\tabletypesize{\small}
	\tablecolumns{6}
	\tablewidth{0pt}
	\tablecaption{A list of X-ray outbursts of radio-loud transient magnetars. \label{tab:radio-loud_outburst}}
	\tablehead{
		\colhead{Source} & 
		\colhead{State} & 
		\colhead{$L_{\rm x}$} &
		\colhead{$L_{\rm sd}$} &	
		\colhead{Distance} &
		\colhead{PF} &		
		\colhead{Energy} &		
		\colhead{References}  
		\\
		\colhead{} & 
		\colhead{} & 
		\colhead{(erg~s$^{-1}$)} & 	
		\colhead{(erg~s$^{-1}$)} & 	
		\colhead{(kpc)} & 			
		\colhead{} & 			
		\colhead{(keV)} & 
	}
	\startdata	
1E~1547.0$-$5408 & Quiescent & $2.2\times10^{33}$ & $2.11\times10^{35}$ & 4.5 & 0.205 & 0.5-2 & 1,2 \\ 
& Outburst (2008) & $2.3\times10^{35}$ &  &  & 0.26 & 0.5-10 & 1,3 \\ 
& Outburst (2009) & $5.0\times10^{35}$ &  &  & 0.13 & 0.5-3 & 1,4 \\ 
XTE J1810$-$197 & Quiescent & $2.5\times10^{34}$ & $1.80\times10^{33}$ & 3.5 & 0.212 & 0.5-2 & 1,2 \\ 
& Outburst (2003) & $1.7\times10^{35}$ &  &  & 0.43 & 1-1.5 & 1,5 \\ 
& Outburst (2018) & $2.5\times10^{35}$ &  &  & 0.27 & 0.5-2 & 6,7 \\ 
PSR J1622$-$4950 & Quiescent & $<7.7\times10^{32}$ & $8.27\times10^{33}$ & 9 &  &  & 8 \\ 
& Outburst (2017) & $1.5\times10^{35}$ &  &  & 0.04 & 0.3-6 & 8 \\ 
SGR 1745$-$1900 & Quiescent & $4.7\times10^{33}$ &$1.02\times10^{34}$ & 8.3 & 0.26 & 0.5-7 & 9,2 \\ 
& Outburst (2013) & $6.8\times10^{35}$ &  &  & 0.45 & 0.3-3.5 & 1,10 \\ 
Swift J1818.0$-$1607 & Quiescent & $<1.7\times10^{34}$ & $1.40\times10^{36}$ & 6.5 &  & & 11 \\ 
& Outburst (2020) & $1.9\times10^{35}$ &  &  & 0.52 & 1-3 & 11,12 \\ 
SGR 1935$+$2154 & Quiescent & $1.1\times10^{34}$ & $1.65\times10^{34}$ & 9 & 0.1 & 0.5-2 & 13,2 \\
& Outburst (2014) & $2.5\times10^{34}$ &  &  & 0.17 & 0.5-1.5 & 1,14 \\ 
& Outburst (2020) & $1.6\times10^{34}$ &  &  & 0.14 & 0.7-3 & 13,15 \\ 
PSR J1119$-$6127 & Quiescent & $5.7\times10^{32}$ & $2.33\times10^{36}$ & 8.4 & 0.74 & 0.5-2 & 1,16 \\ 
& Outburst (2016) & $3.7\times10^{35}$ &  &  & 0.67 & 0.7-3 & 1,17 \\ 
	\enddata
	\tablecomments{\\
	$L_{\rm x}$: Observed X-ray Luminosity (0.3-10 keV) assuming the distance in the right column. \\
	$L_{\rm sd}$: Spin-down luminosity.\\
	PF: X-ray pulsed fraction defined in the energy band in the right column. \\
	References: 
	1. \citet{2018MNRAS.474..961C}; 2. \citet{2019MNRAS.485.4274H}; 3. \citet{2010MNRAS.408.1387I};  4. \citet{2011AA...529A..19B}
	5. \citet{2004ApJ...605..368G}; 6. \citet{2020arXiv200508410P}; 7. \citet{2021MNRAS.504.5244B}; 8. \citet{2018ApJ...856..180C}; 9. \citet{2020ApJ...894..159R}; 10. \citet{2015MNRAS.449.2685C}; 11. \citet{2020ApJ...902....1H}; 12. \citet{2020ApJ...896L..30E}; 13. \citet{2020ApJ...902L...2B}; 14. \citet{2016MNRAS.457.3448I}; 15. \citet{2020ApJ...905L..31G}; 16. \citet{2005ApJ...630..489G}; 17. \citet{2018ApJ...869..180A}}
\end{deluxetable*}

\begin{deluxetable*}{llrrrrrr}
	\tabletypesize{\small}
	\tablecolumns{6}
	\tablewidth{0pt}
	\tablecaption{A list of X-ray outbursts of radio-quiet transient magnetars. \label{tab:radio-quiet_outburst}}
	\tablehead{
		\colhead{Source} & 
		\colhead{State} & 
		\colhead{$L_{\rm x}$} &
		\colhead{$L_{\rm sd}$} &	
		\colhead{Distance} &
		\colhead{PF} &		
		\colhead{Energy} &		
		\colhead{References}  
		\\
		\colhead{} & 
		\colhead{} & 
		\colhead{(erg~s$^{-1}$)} & 	
		\colhead{(erg~s$^{-1}$)} & 	
		\colhead{(kpc)} & 			
		\colhead{} & 			
		\colhead{(keV)} & 
	}
	\startdata
\src & Outburst (2021) & XX  & $1.82\times10^{34}$ & 10 & 0.15 & 1-2 & this work\\	
SGR 0418$+$5729 & Quiescent & $7.0\times10^{30}$ & $2.11\times10^{29}$ & 2 & 0.37 & 0.5-2 & 1,2 \\ 
& Outburst (2009) & $1.6\times10^{34}$ &  &  & 0.42 & 1.5-2.5 & 1,3 \\ 
SGR 0501$+$4516 & Quiescent & $1.2\times10^{33}$ & $1.22\times10^{33}$ & 1.5 & 0.28 & 0.5-2 & 1,2 \\ 
& Outburst (2008) & $3.4\times10^{34}$ &  &  & 0.24 & 0.3-2 & 1,4 \\ 
1E 1048.1$-$5937 & Quiescent & $8.6\times10^{34}$ & $3.29\times10^{33}$ & 9 & 0.584 & 0.5-2 & 1,2 \\ 
& Outburst (2011) & $5.7\times10^{35}$ &  &  & 0.10 & 1-10 & 1,5 \\ 
& Outburst (2016) & $3.7\times10^{35}$ &  &  & 0.51 & 3-7 & 1,6 \\ 
CXOU J164710.2$-$455216 & Quiescent & $3.3\times10^{33}$ & $1.32\times10^{31}$ & 4 & 0.47 & 0.5-2 & 1,2 \\ 
& Outburst (2006) & $1.2\times10^{35}$ &  &  & 0.10 & 0.5-4 & 1,7 \\ 
& Outburst (2011) & $2.1\times10^{34}$ &  &  & 0.60 & 0.5-4 & 1,7 \\ 
& Outburst (2017) & $1.9\times10^{34}$ &  &  & 0.60 & 0.3-2.5 & 8 \\ 
& Outburst (2018) & $8.0\times10^{34}$ &  &  & 0.45 & 0.3-2.5 & 8 \\ 
SGR 1806$-$20 & Quiescent & $8.2\times10^{34}$ & $4.54\times10^{34}$ & 8.7 & 0.1 & 0.5-4 & 1,2 \\ 
& Outburst (2004) & $3.6\times10^{35}$ &  &  & 0.03 & 2-10 & 1,9 \\ 
Swift J1822.3$-$1606 & Quiescent & $2.0\times10^{32}$ & $1.38\times10^{30}$ & 1.6 & 0.33 & 0.5-2 & 1,2 \\ 
& Outburst (2011) & $8.0\times10^{34}$ &  &  & 0.43 & 2-8 & 1,10 \\ 
1E 2259$+$586 & Quiescent & $5.8\times10^{34}$ & $5.61\times10^{31}$ & 3.2 & 0.233 & 0.5-2 & 1,2 \\ 
& Outburst (2002) & $1.2\times10^{35}$ &  &  & 0.322 & 0.1-2 & 1,11 \\ 
SGR 1627$-$41 & Quiescent & $1.2\times10^{33}$ & $4.29\times10^{34}$ & 11 &  &  & 1 \\ 
& Outburst (1998) & $5.2\times10^{34}$ &  &  & 0.10 & 0.1-10 & 1,12 \\ 
& Outburst (2008) & $3.2\times10^{35}$ &  &  & 0.13 & 2-10 & 1,13 \\ 
SGR 1833$-$0832 & Quiescent & $<8.0\times10^{33}$ & $3.18\times10^{32}$ & 10 &  &  & 1 \\ 
& Outburst (2010) & $1.0\times10^{35}$ &  &  & 0.34 & 0.2-4 & 1,14 \\ 
Swift J1834.9$-$0846 & Quiescent & $<2.0\times10^{32}$ & $2.05\times10^{34}$ & 4.2 &  &  & 1 \\ 
& Outburst (2011) & $1.0\times10^{35}$ &  &  & 0.85 & 2-10 & 1,15 \\ 
SGR 1830$-$0645 & Quiescent & $<2.0\times10^{34}$ & $2.44\times10^{32}$ & 10 &  &  & 16 \\ 
& Outburst (2020) & $6.0\times10^{35}$ &  &  & 0.63 & 0.3-2 & 16 \\ 
SGR 1900$+$14 & Quiescent & $1.3\times10^{35}$ & $2.58\times10^{34}$ & 12.5 & 0.11 & 0.5-2 & 1,2 \\ 
& Outburst (2001) & $3.5\times10^{35}$ &  &  & 0.10 & 0.8-6.5 & 1,17 \\ 
& Outburst (2006) & $2.4\times10^{35}$ &  &  & 0.151 & 0.8-4 & 1,18 \\ 
1E 1841$-$045 & Quiescent & $4.3\times10^{35}$ & $9.84\times10^{32}$ & 8.5 & 0.11 & 0.5-2 & 1,2 \\ 
& Outburst (2011) & $1.7\times10^{36}$ &  &  & 0.10 & 0.5-2 & 1,19 \\ 
4U 0142$+$61 & Quiescent & $3.6\times10^{35}$ & $1.21\times10^{32}$ & 3.6 & 0.047 & 0.5-2 & 1,2 \\ 
& Outburst (2011) & $1.2\times10^{36}$ &  &  & 0.17 & 0.7-10 & 1,20 \\ 
& Outburst (2015) & $1.3\times10^{36}$ &  &  & 0.09 & 0.7-10 & 1,20 
	\enddata
	\tablecomments{\\
	Definitions of the columns are the same as Table~\ref{tab:radio-loud_outburst}.\\
	References: 
	1.\citet{2018MNRAS.474..961C}; 2.\citet{2019MNRAS.485.4274H}; 3.\citet{2010MNRAS.405.1787E}; 4.\citet{2010ApJ...722..899G}; 5.\citet{2015ApJ...800...33A}; 6.\citet{2020ApJ...889..160A}; 7.\citet{2014MNRAS.441.1305R}; 8.\citet{2019MNRAS.484.2931B}; 9.\citet{2007ApJ...654..470W}; 10.\citet{2011ApJ...743L..38L}; 11.\citet{2008ApJ...686..520Z}; 12.\citet{1999ApJ...519L.139W}; 13.\citet{2009MNRAS.399L..44E}; 14.\citet{2010ApJ...718..331G}; 15.\citet{2012ApJ...748...26K}; 16.\citet{2021ApJ...907L..34C}; 17.\citet{2011ApJ...728..160G}; 18.\citet{2006ApJ...653.1423M}; 19.\citet{2013ApJ...779..163A}; 20.\citet{2017ApJ...834..163A}
	}
\end{deluxetable*}




  

\bibliographystyle{aasjournal}
\bibliography{reference}



\end{document}